\documentclass[iop,apj,tighten]{emulateapj}

\newcommand{\lya}{Ly$\alpha$}

\newcommand{\hi}{\ion{H}{1}}
\newcommand{\hii}{\ion{H}{2}}
\newcommand{\heii}{\ion{He}{2}}
\newcommand{\oi}{\ion{O}{1}}
\newcommand{\cii}{\ion{C}{2}}

\newcommand{\civ}{\ion{C}{4}}
\newcommand{\siii}{\ion{Si}{2}}

\newcommand{\siiv}{\ion{Si}{4}}
\newcommand{\mgi}{\ion{Mg}{1}}
\newcommand{\mgii}{\ion{Mg}{2}}
\newcommand{\feii}{\ion{Fe}{2}}

\newcommand{\moi}{{\rm O \; \mbox{\tiny I}}}
\newcommand{\msiii}{{\rm Si \; \mbox{\tiny II}}}

\newcommand{\msiiv}{{\rm Si \; \mbox{\tiny IV}}}
\newcommand{\mcii}{{\rm C \; \mbox{\tiny II}}}
\newcommand{\mciv}{{\rm C \; \mbox{\tiny IV}}}
\newcommand{\kms}{km~s$^{-1}$}

\newcommand{\dv}{$\Delta v_{90}$}
\newcommand{\W}{$W_{1526}$}
\newcommand{\ew}{$W_{0}$}
\newcommand{\Msun}{M$_\odot$}

\shorttitle{Low-Ionization Absorption Lines at $z \sim 6$} 

\shortauthors{Becker et al.}

\begin{document}

\title{High-Redshift Metals. II. Probing Reionization Galaxies with Low-Ionization Absorption Lines at Redshift Six\altaffilmark{1,2}}

\author{George D. Becker\altaffilmark{3}, Wallace L. W. Sargent\altaffilmark{4}, Michael Rauch\altaffilmark{5}, Alexander P. Calverley\altaffilmark{3}}

\altaffiltext{1}{The observations were made at the W.M. Keck Observatory which is operated as a scientific partnership between the California Institute of Technology and the University of California; it was made possible by the generous support of the W.M. Keck Foundation.}
\altaffiltext{2}{This paper includes data gathered with the 6.5 meter Magellan Telescopes located at Las Campanas Observatory, Chile.}
\altaffiltext{3}{Kavli Institute for Cosmology and Institute
  of Astronomy, Madingley Road, Cambridge, CB3 0HA, UK;
  gdb@ast.cam.ac.uk, acalver@ast.cam.ac.uk}
\altaffiltext{4}{Palomar Observatory, California Institute of
  Technology, Pasadena, CA 91125, USA; wws@astro.caltech.edu}
\altaffiltext{5}{Carnegie Observatories, 813 Santa Barbara Street,
  Pasadena, CA 91101, USA; mr@obs.carnegiescience.edu}

\slugcomment{Submitted to ApJ}

\begin{abstract}

We present a survey for low-ionization metal absorption line systems towards 17 QSOs at redshifts $z_{\rm em} = 5.8-6.4$.  Nine of our objects were observed at high resolution with either Keck/HIRES or Magellan/MIKE, and the remainder at moderate resolution with Keck/ESI.  The survey spans $5.3 < z_{\rm abs} < 6.4$ and has a pathlength interval $\Delta X = 39.5$, or $\Delta z = 8.0$.  In total we detect ten systems, five of which are new discoveries.   The line-of-sight number density, $\ell(X) = 0.25^{+0.21}_{-0.13}$  (95\% confidence), is consistent with the combined number density at $z \sim 3$ of DLAs and sub-DLAs, which comprise the main population of low-ionization systems at lower redshifts.  This apparent lack of evolution may occur because low ionization systems are hosted by lower-mass halos at higher redshifts, or because the mean cross section of low-ionization gas at a given halo mass increases with redshift due to the higher densities and lower ionizing background.  The roughly  constant number density notably contrasts with the sharp decline at $z > 5.3$ in the number density of highly-ionized systems traced by \civ.  The low-ionization systems at $z \sim 6$ span a similar range of velocity widths as lower-redshift sub-DLAs but have significantly weaker lines at a given width.  This implies that the mass-metallicity relation of the host galaxies evolves towards lower metallicities at higher redshifts.  These systems lack strong \siiv\ and \civ, which are common among lower-redshift DLAs and sub-DLAs.  This is consistent, however, with a similar decrease in the metallicity of the low- and high-ionization phases, and does not necessarily indicate a lack of nearby, highly-ionized gas.  The high number density of low-ionization systems at $z \sim 6$ suggests that we may be detecting galaxies below the current limits of $i$-dropout and \lya\ emission galaxy surveys.  These systems may therefore be the first direct probes of the `typical' galaxies responsible for hydrogen reionization.

\end{abstract}

\keywords{cosmology: observations --- cosmology: early universe ---
  intergalactic medium --- quasars: absorption lines}

\section{Introduction}\label{sec:intro}

The universe at redshift six remains one of the most challenging observational frontiers.  The high opacity of the \lya\ forest inhibits detailed measurements of the intergalactic medium (IGM), which hosts the vast majority of baryons in the early universe.  At the same time, extreme luminosity distances make it difficult to assemble large samples of galaxies, or to study individual objects in detail.  Despite these challenges, however, significant observational progress is being made at $z \sim 6$.  Statistics of the scarce transmitted flux in the \lya\ forest are being combined with insights from quasar proximity zones to learn about the very high-redshift IGM \citep[e.g.,][]{fan2006b,lidz2006,becker2007,gallerani2008,maselli2009,bolton2010,calverley2010,carilli2010,mesinger2010}.  Deep galaxy surveys from the ground and from space are also beginning to yield considerable information on the properties of galaxies at $z \sim 6$ and beyond \citep[e.g.,][]{bouwens2007,bouwens2010,stanway2007,martin2008,ouchi2008,ouchi2009,bunker2010,hu2010,labbe2010,mclure2010,oesch2010,stark2010}.  A more complete understanding of the universe at such early times, however, will require insight from a wide variety of observations.

Metal absorption lines offer a unique probe into the high-redshift universe.  Transitions with rest wavelengths longer than \hi\ \lya\ will remain visible in QSO and GRB spectra even when the forest becomes saturated.  While absorption lines studies at $z \sim 6$ are still in their early days, metal lines are expected to provide insight into past and ongoing star formation, the action of galactic winds, as well as conditions in the interstellar media of  galaxies, just as they do at lower redshifts \citep[e.g.,][]{songaila2001,simcoe2002,schaye2003,adelberger2005,wolfe2005,scannapieco2006,fox2007a}.  At high redshifts, the role of metal absorption lines in studying galaxy evolution becomes increasingly significant as galaxies become more difficult to study in emission. 

Metal lines play another unique role at $z \gtrsim 6$ as potential probes of hydrogen reionization.  In a scenario where galaxies provide most of the ionizing photons, dense regions of the IGM may become chemically enriched early on, yet remain significantly neutral until the end of reionization due to the short recombination times at high densities.  High-density peaks are commonly predicted to host the last pockets of neutral gas following the growth and overlap of large \hii\ regions \citep{me2000, ciardi2003, furoh2005, gnedinfan2006}. If these regions are metal-enriched, then ÒforestsÓ of low-ionization absorption lines such as \oi, \siii, and \cii\ may be observable during this last phase of reionization \citep{oh2002, furloeb2003}.

We have conducted a two-part survey for metals out to $z \sim 6$.  In the first paper of this series \citep[][hereafter Paper I]{becker2009} we used high-resolution, near-infrared spectra to search for \civ\ over $z = 5.3-6.0$.  \civ\ is a widely-used tracer of highly-ionized metals in the IGM at $z < 5$ \citep[e.g.,][]{songaila2001}, and initial studies had indicated that \civ\ may remain abundant out to $z \sim 6$ \citep{simcoe2006b, rw2006}.  In contrast, the longer path length and greater sensitivity of our data allowed us to demonstrate that the number density of \civ\ systems declines by at least a factor of four from $z \sim 4.5$ to $z > 5.3$.  This downturn was also seen by \citet{rw2009} using a larger sample of low-resolution spectra.  Since \civ\ is expected to be an effective tracer of metals in the IGM at $z \sim 5-6$ \citep{oppenheimer2006, oppenheimer2009}, the decline suggests that there are fewer metals in the IGM towards higher redshifts, consistent with gradual enrichment of the IGM by star-forming galaxies.  The apparent rapid evolution of \civ\ near $z \sim 5$ may also point to changes in the ionization state of the metal-enriched gas, perhaps due to a hardening of the UV background associated with the beginning of \heii\ reionization \citep{madauhaardt2009}.

In this paper we present a complementary search for low-ionization metal absorbers over $5.3 < z < 6.4$.  This is an extension of an earlier work \citep{becker2006}, in which we used high-resolutions spectra of nine QSOs spanning $4.9 < z_{\rm em} < 6.4$, including five at $z_{\rm em} > 5.8$, to search for `\oi\ absorbers,' so named due to the presence of \oi~$\lambda1302$.  Although we identified a possible excess of these systems at $z \gtrsim 6$, the significance of that result remained unclear due to the relatively short survey pathlength, and the fact that the $z \sim 6$ systems occurred almost entirely along a single sightline.  Here we increase our sample of $z_{\rm em} \sim 6$ QSOs to 17.  The expanded survey offers us the opportunity to gather a more representative sample of low-ionization systems, providing insight into early galaxies and possibly the tail end of hydrogen reionization.

The remainder of the paper is organized as follows.  The data are described in Section~\ref{sec:data}.  We present our detections and the measurements for individual systems in Section~\ref{sec:systems}.  In Section~\ref{sec:sample} we analyze a number of ensemble properties, including the number density, ionic mass densities, relative abundance patterns, lack of strong high-ionization lines, and velocity width distribution.  We also present evidence that the mass-metallicity relation of the host galaxies of low-ionization absorbers evolves out to $z \sim 6$.  Possible causes for the apparent lack of evolution in the number density of low-ionization absorbers are discussed in Section~\ref{sec:discussion}, where we also consider the likely hosts of these systems.  Finally, we summarize our conclusions in Section~\ref{sec:summary}.  Throughout this paper we assume $(\Omega_{\rm m}, \Omega_{\Lambda}, h) = (0.274, 0.726, 0.705)$ \citep{komatsu2009}.

\section{The Data}\label{sec:data}

\begin{deluxetable*}{lcccc}
   \tabletypesize{} 
   \tablewidth{4.0in}
   \centering
   \tablecolumns{5}
   \tablecaption{Data} 
   \tablehead{\colhead{QSO} & 
              \colhead{$z_{\rm em}$\tablenotemark{a}} &
              \colhead{Instrument} &
	      \colhead{$t_{\rm exp}$} &
              \colhead{$S/N_{\rm res}$\tablenotemark{b}} \\
               & & & (hrs) & } 
   \startdata
   \multicolumn{5}{c}{High Resolution} \\
   SDSS~J1044$-$0125  &  5.782  &  MIKE   &   7.5  &  10.7 -- 19.7 \\
   SDSS~J0836$+$0054  &  5.81   &  HIRES  &  12.5  &  15.0 -- 23.4 \\
   SDSS~J0002$+$2550  &  5.82   &  HIRES  &  14.2  &  13.8 -- 28.0 \\
   SDSS~J0818$+$1722  &  6.00   &  HIRES  &   8.3  &  11.8 -- 18.5 \\
   SDSS~J1306$+$0356  &  6.016  &  MIKE   &   6.7  & \ 7.3 -- 15.9 \\
   SDSS~J1048$+$4637  &  6.23   &  HIRES  &   5.0  & \ 7.2 -- 12.9 \\
   SDSS~J1623$+$3112  &  6.247  &  HIRES  &  12.5  & \ 6.0 -- 20.6 \\
   SDSS~J1030$+$0524  &  6.308  &  HIRES  &  10.0  & \ 6.7 -- 16.9 \\
   SDSS~J1148$+$5251  &  6.419  &  HIRES  &  14.2  &  10.5 -- 21.6 \\
   \multicolumn{5}{c}{Moderate Resolution} \\
   SDSS~J0005$-$0006  &  5.85   &  ESI    &   4.2  &  27.6 -- 50.0 \\
   SDSS~J0203$+$0012  &  5.85   &  ESI    &   1.8  &  14.0 -- 22.8 \\
   ULAS~J0148$+$0600  &  6.00   &  ESI    &   3.1  &  50.9 -- 85.9 \\
   SDSS~J1630$+$4012  &  6.05   &  ESI    &   0.6  &  14.7 -- 27.7 \\
   SDSS~J0353$+$0104  &  6.05   &  ESI    &   3.7  &  27.8 -- 55.4 \\
   SDSS~J2054$-$0005  &  6.06   &  ESI    &   3.3  &  33.3 -- 50.8 \\
   SDSS~J2315$-$0023  &  6.12   &  ESI    &   6.7  &  27.5 -- 70.6 \\
   CFHQS~J0050$+$3445 &  6.25   &  ESI    &   1.7  &  21.7 -- 55.4    
   \enddata 
   \tablenotetext{a}{QSO redshifts quoted to three decimal places are
     \mgii\ or CO redshifts from \citet{carilli2010}.  Others are
     discovery redshifts based on \lya+\ion{N}{5} emission, or are
     measured from the apparent start of the \lya\ forest.}
   \tablenotetext{b}{Interquartile values of signal-to-noise in the
     continuum per resolution element within the wavelength interval
     used to search for O\,I.  The resolution elements for HIRES, MIKE, and
     ESI are 6.7, 13.6, and 50 \kms, respectively.}
   \label{tab:obs}
\end{deluxetable*}

We have obtained high- or moderate-resolution spectra of 17 QSOs with emission redshifts $z_{\rm em} = 5.8-6.4$.  This sample contains more than three times the number of $z \ge 5.8$ QSO sight lines than were included in \citet{becker2006}.  Nine of our objects (generally the brightest) were observed at high resolution with either the HIRES spectrograph \citep{vogt1994} on Keck or the MIKE spectrograph \citep{bernstein2002} on Magellan.  This sample includes the five $z \ge 5.8$ QSOs from \citet{becker2006}, although additional data have since been acquired for some of these.  We complement our high resolution sample with a set of eight QSO spectra taken at moderate resolution with ESI \citep{sheinis2002} on Keck.  The data are summarized in Table~\ref{tab:obs}.

All data were reduced using custom sets of IDL routines that included optimal sky subtraction algorithms \citep{kelson2003}.  In order to efficiently reject cosmic rays and bad pixels, a single 1-D spectrum was optimally extracted \citep{horne1986} from all 2-D sky-subtracted frames for a given object simultaneously.  Prior to extraction, relative flux calibration across the orders was performed using response functions generated from standard stars.  Continua redward of the \lya\ forest were fit by hand using a cubic spline.  

The HIRES spectra were taken using the upgraded detector.  Two grating settings with the red cross disperser provided continuous spectral coverage up to nearly $1~\micron$.  The spectra were taken using a 0\farcs86 slit, which provides a resolution FWHM of 6.7\,\kms.  The final 1D spectra were binned using 2.1\,\kms\ pixels.  Telluric absorption corrections based on a standard star were also applied to the individual 2-D frames prior to final extraction. For objects observed at multiple times during the year, changes in heliocentric velocity allowed telluric features to be efficiently excluded from the final spectra.  Data from \citet{becker2006} were re-reduced to take advantage of the improvements in the reduction pipeline.  Only data taken with the upgraded detector were included.

MIKE spectra were taken using a single setting that gave continuous coverage up to 9400~\AA.  We used a 1\farcs00 slit, which gives 13.6\,\kms\ resolution in the red.  The 1D spectra were binned using 5.0\,\kms\ pixels.  Telluric corrections were performed after extracting the 1D spectra using a high-resolution, high signal-to-noise transmission spectrum derived from solar observations \citep{hinkle2003}, which was shifted, smoothed, and scaled in optical depth to match the QSO spectra.

ESI provides complete wavelength coverage up to 1~$\micron$ in a single setting.  We used both the 0\farcs75 and 1\farcs00 slits, depending on the seeing. Our combined resolution was 50\,\kms, and was roughly consistent across all objects.  The 1D spectra were binned using 15.0\,\kms\ pixels.  Telluric corrections were performed after extraction using the transmission spectrum from \citet{hinkle2003}.

Finally, we include Keck NIRSPEC echelle spectra (FWHM = 23\,\kms) covering high-ionization lines (\siiv\ and \civ) for SDSS~J0818$+$1722 and SDSS~J1148$+$5251.  These data are described in Paper I.

\section{Detected Systems}\label{sec:systems}

\begin{figure}
   \epsscale{0.90} 
   \centering 
   \plotone{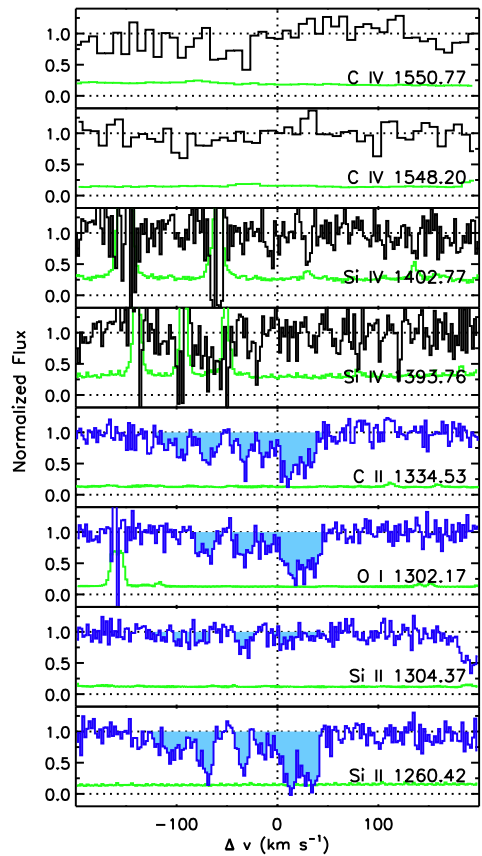}
   \caption{Stacked velocity plot for the absorption systems at $z=5.7911$ towards SDSS~J0818$+$1722.  Data covering the low-ionization lines and \siiv\ are from HIRES, while the data for \civ\ are from NIRSPEC.  Histograms show the normalized flux, and the formal error array is plotted along the bottom of each panel.  Shaded regions indicate detected transitions.  
     \label{fig:0818_z5.7911}}
\end{figure}

\begin{figure}
   \epsscale{0.90} 
   \centering 
   \plotone{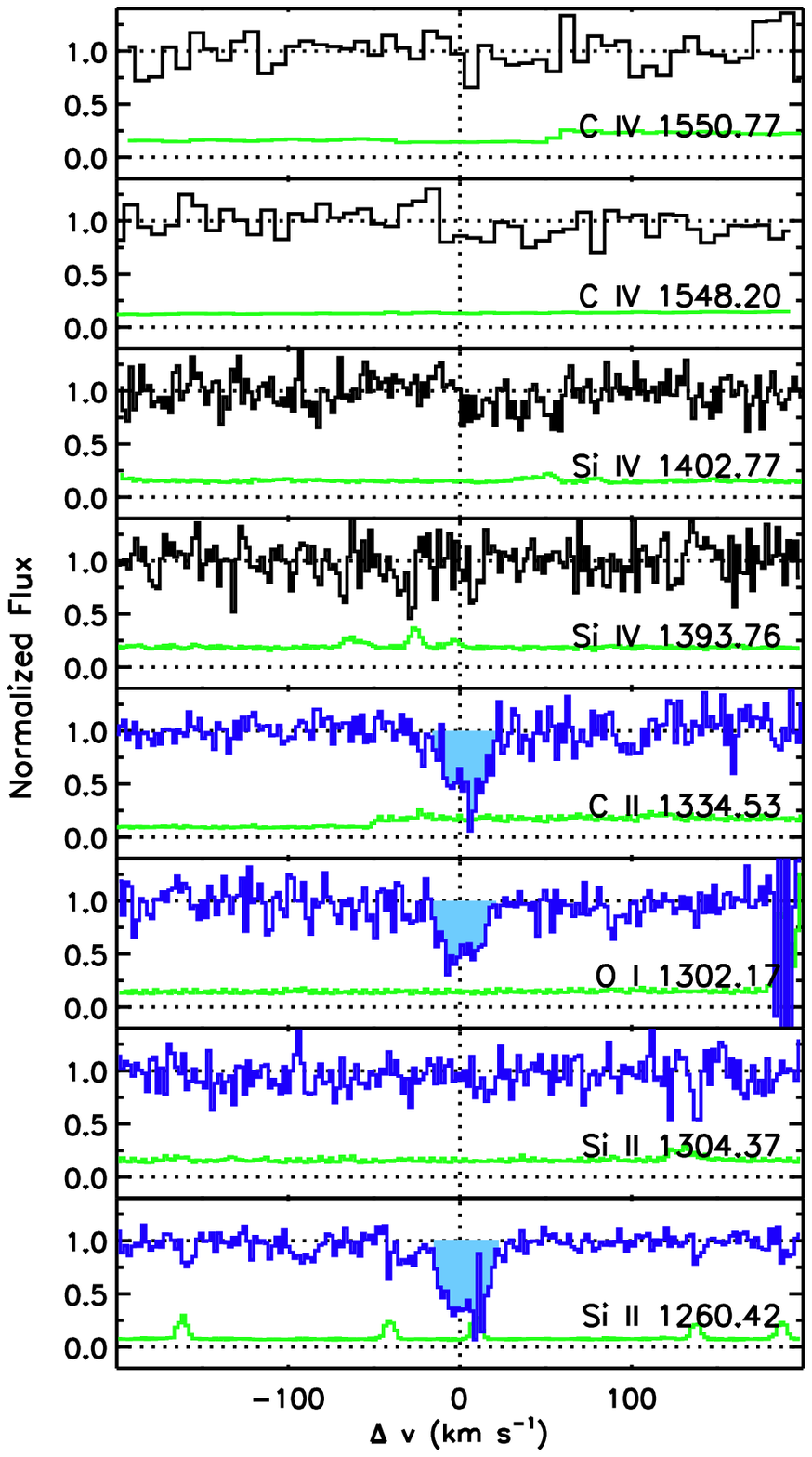}
   \caption{Stacked velocity plot for the absorption systems at $z=5.8765$ towards SDSS~J0818$+$1722.  Data covering the low-ionization lines and \siiv\ are from HIRES, while the data for \civ\ are from NIRSPEC.  Lines are as in Figure~\ref{fig:0818_z5.7911}.  \siii~$\lambda1260$ is partially affected by a skyline residual.
     \label{fig:0818_z5.8765}}
\end{figure}

\begin{figure}
   \epsscale{0.90} 
   \centering 
   \plotone{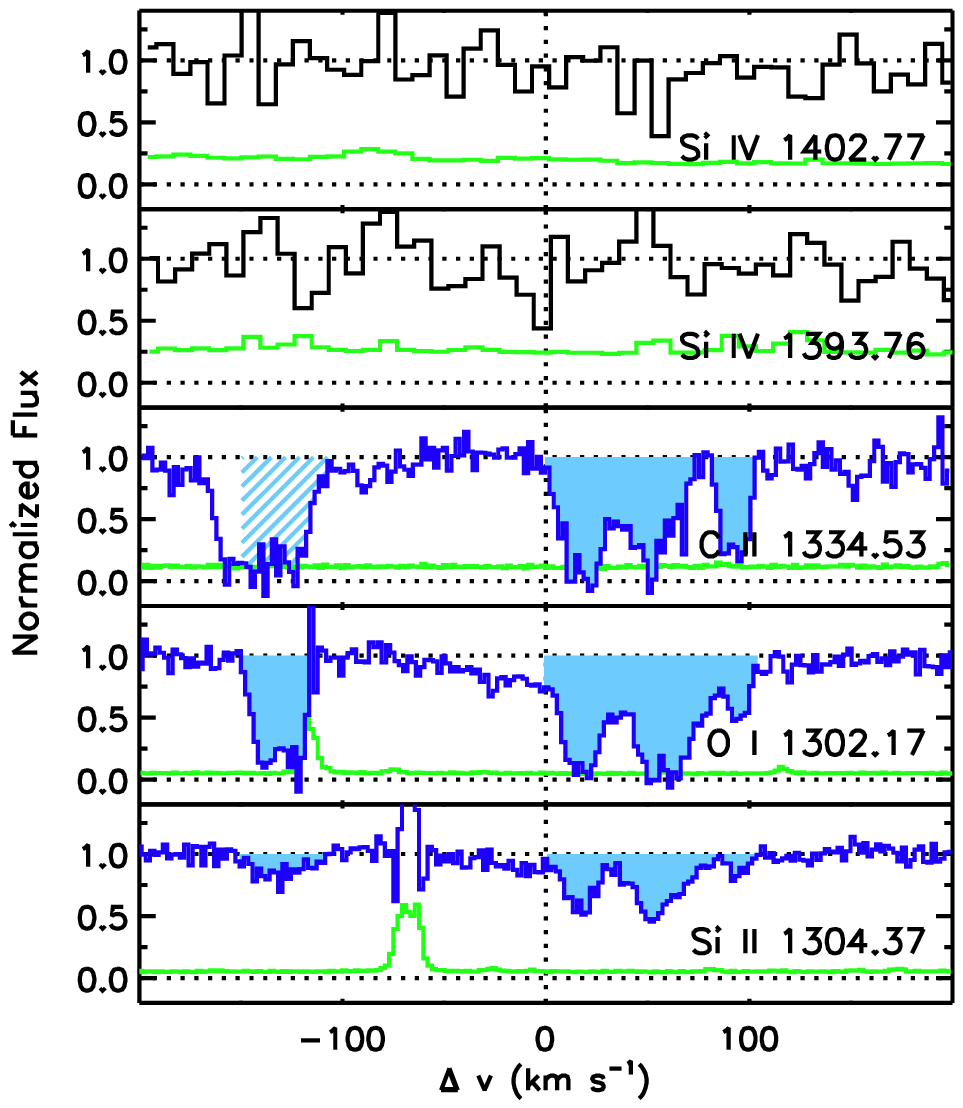}
   \caption{Stacked velocity plot for the absorption systems at $z=5.8415$ towards SDSS~J1623$+$3112 \citep{becker2006}.  All data are from HIRES, although the the data covering \siiv\ are binned to 8.4\,\kms\ pixels for display.  Lines are as in Figure~\ref{fig:0818_z5.7911}.  \cii\ in the bluest component is blended with a strong \mgi\ line at lower redshift.  \oi\ and \siii~$\lambda1304$ are both partially blended with a \civ\ system at $z=4.754$, which is visible over $-50~{\rm km\,s}^{-1} < \Delta v < 0~{\rm km\,s}^{-1}$.
     \label{fig:1623_z5.8415}}
\end{figure}

\begin{figure}
   \epsscale{0.90} 
   \centering 
   \plotone{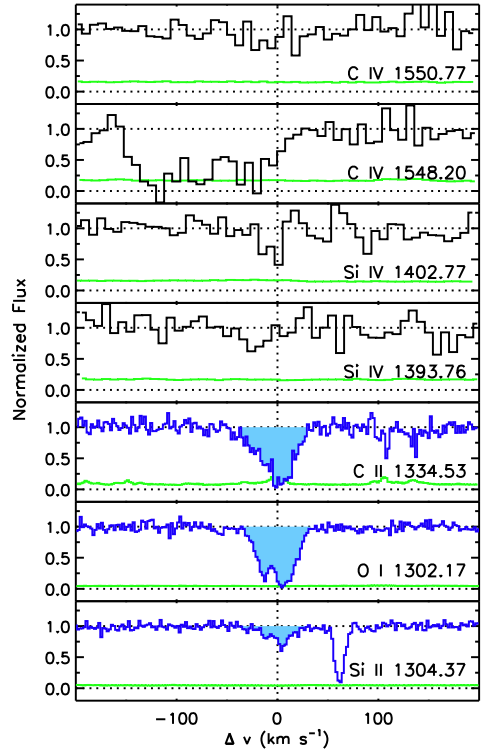}
   \caption{Stacked velocity plot for the absorption systems at $z=6.0115$ towards SDSS~J1148$+$5251 \citep{becker2006}.  Data covering the low-ionization lines are from HIRES, while  the data for \siiv\ and \civ\ are from NIRSPEC.  Lines are as in Figure~\ref{fig:0818_z5.7911}.  The absorption covering the blue half of the \civ~$\lambda1548$ region is unrelated \feii\ at lower redshift. 
     \label{fig:1148_z6.0115}}
\end{figure}

\begin{figure}
   \epsscale{0.90} 
   \centering 
   \plotone{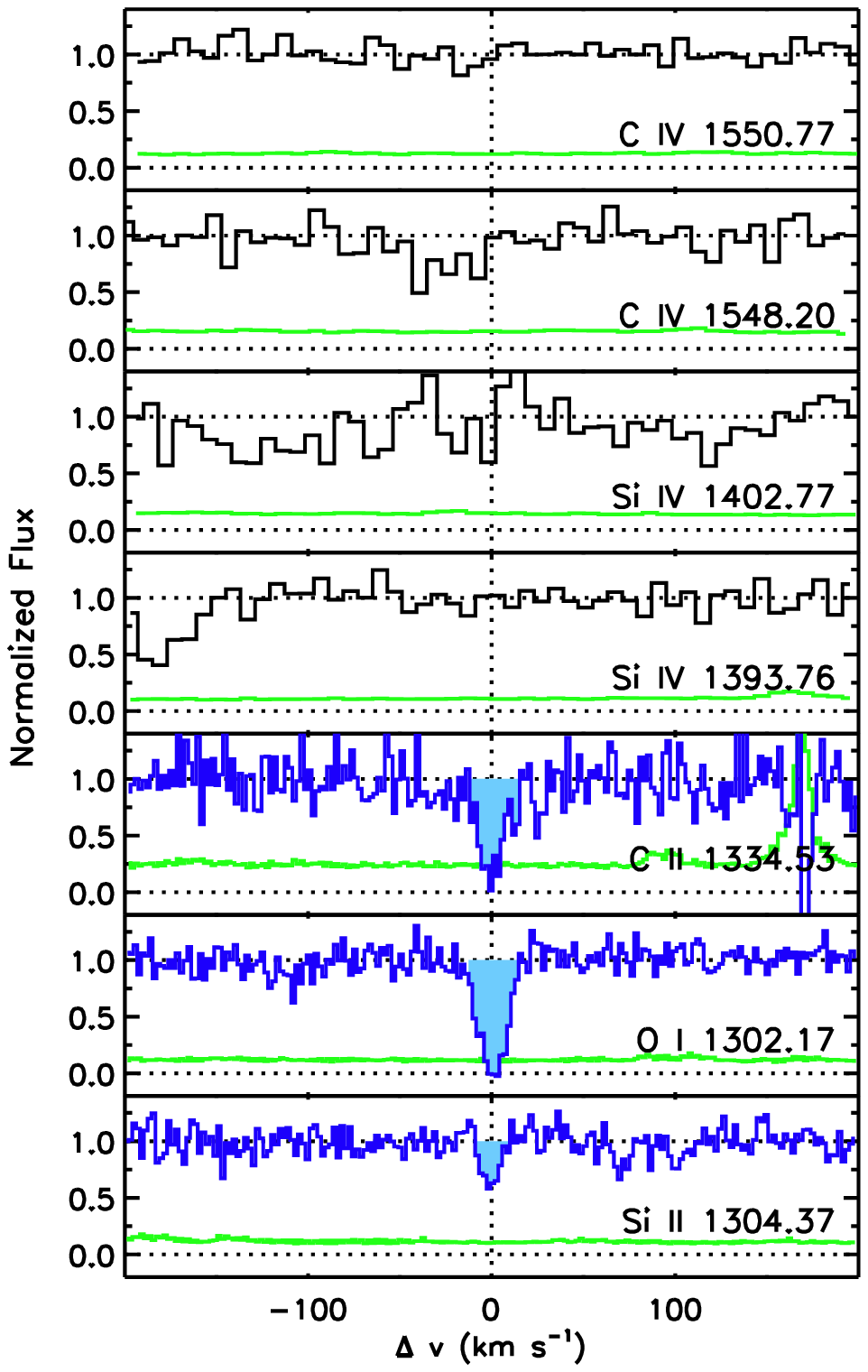}
   \caption{Stacked velocity plot for the absorption systems at $z=6.1312$ towards SDSS~J1148$+$5251 \citep{becker2006}.  Data covering the low-ionization lines are from HIRES, while  the data for \siiv\ and \civ\ are from NIRSPEC.  Lines are as in Figure~\ref{fig:0818_z5.7911}.
     \label{fig:1148_z6.1312}}
\end{figure}

\begin{figure}
   \epsscale{0.90} 
   \centering 
   \plotone{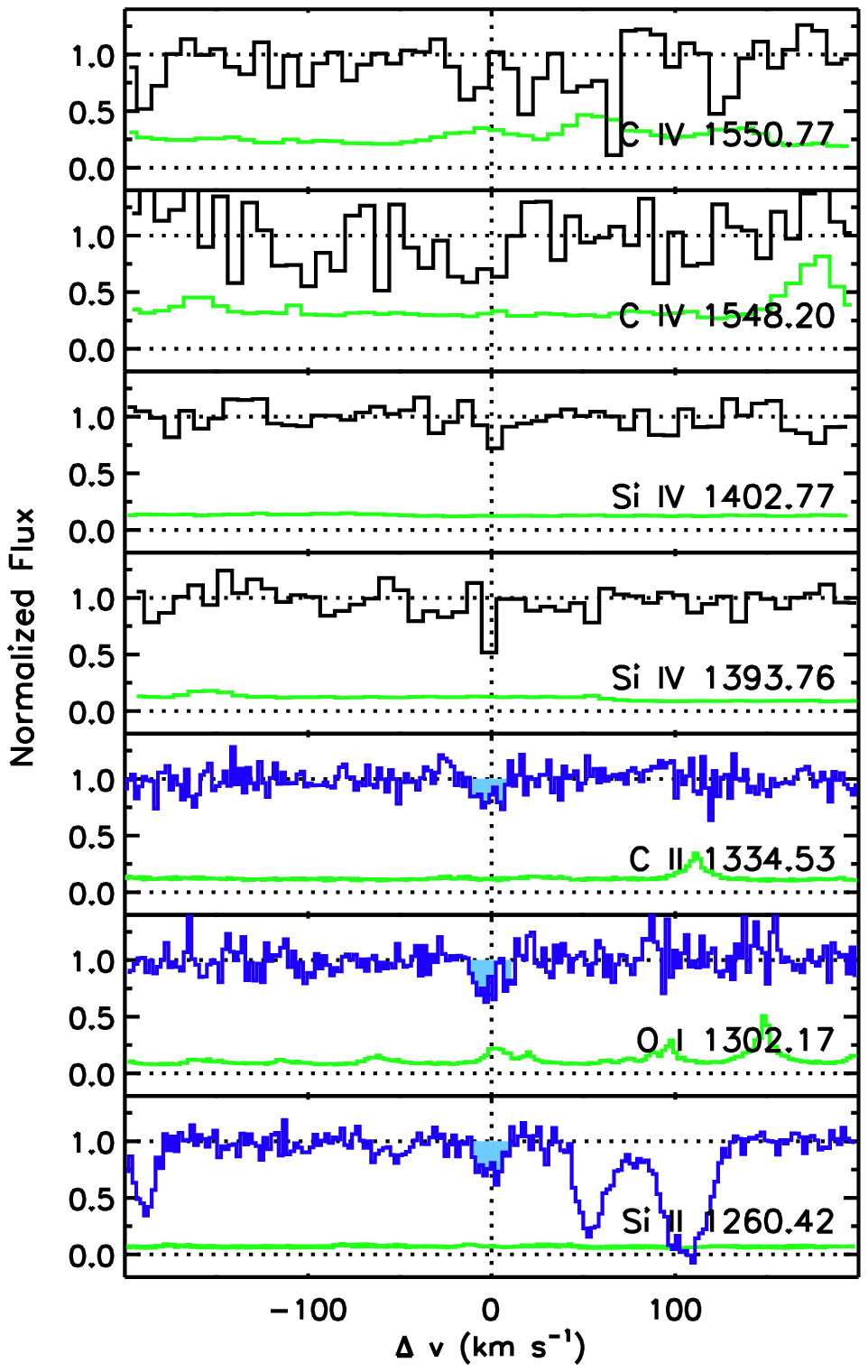}
   \caption{Stacked velocity plot for the absorption systems at $z=6.1988$ towards SDSS~J1148$+$5251 \citep{becker2006}.  Data covering the low-ionization lines are from HIRES, while  the data for \siiv\ and \civ\ are from NIRSPEC.  Lines are as in Figure~\ref{fig:0818_z5.7911}.
     \label{fig:1148_z6.1988}}
\end{figure}

\begin{figure}
   \epsscale{0.90} 
   \centering 
   \plotone{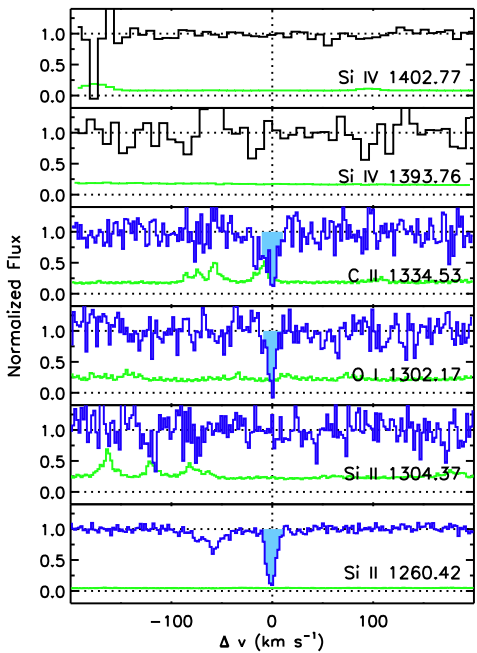}
   \caption{Stacked velocity plot for the absorption systems at $z=6.2575$ towards SDSS~J1148$+$5251 \citep{becker2006}.  Data covering the low-ionization lines are from HIRES, while  the data for \siiv\ are from NIRSPEC.  Lines are as in Figure~\ref{fig:0818_z5.7911}.  \cii\ is partially affected by skyline residuals.
     \label{fig:1148_z6.2575}}
\end{figure}

\begin{figure}
   \epsscale{0.90} 
   \centering 
   \plotone{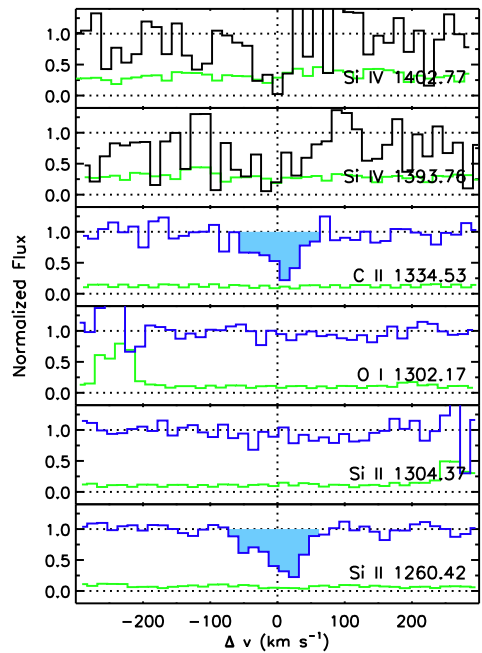}
   \caption{Stacked velocity plot for the absorption systems at $z=5.8865$ towards SDSS~J1630$+$4012.  The data are from ESI.  Lines are as in Figure~\ref{fig:0818_z5.7911}.  Note the change in velocity scale from Figures~\ref{fig:0818_z5.7911}-\ref{fig:1148_z6.2575}.
     \label{fig:1630_z5.8865}}
\end{figure}

\begin{figure}
   \epsscale{0.90} 
   \centering 
   \plotone{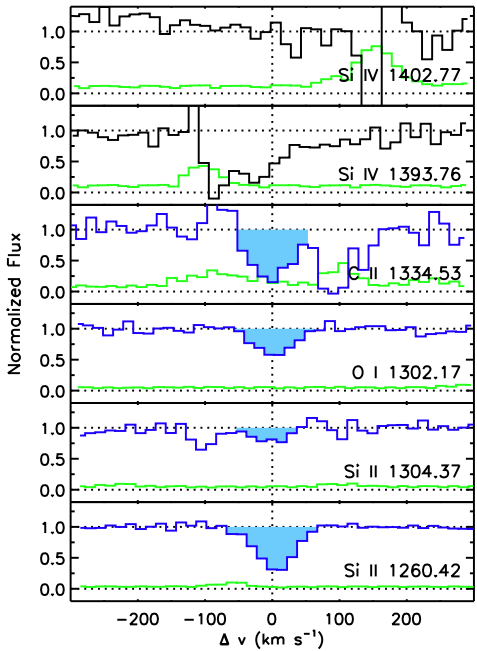}
   \caption{Stacked velocity plot for the absorption systems at $z=5.9776$ towards SDSS~J2054$-$0005.  The data are from ESI.  Lines are as in Figure~\ref{fig:0818_z5.7911}.  Note the change in velocity scale from Figures~\ref{fig:0818_z5.7911}-\ref{fig:1148_z6.2575}.
     \label{fig:2054_z5.9776}}
\end{figure}

\begin{figure}
   \epsscale{0.90} 
   \centering 
   \plotone{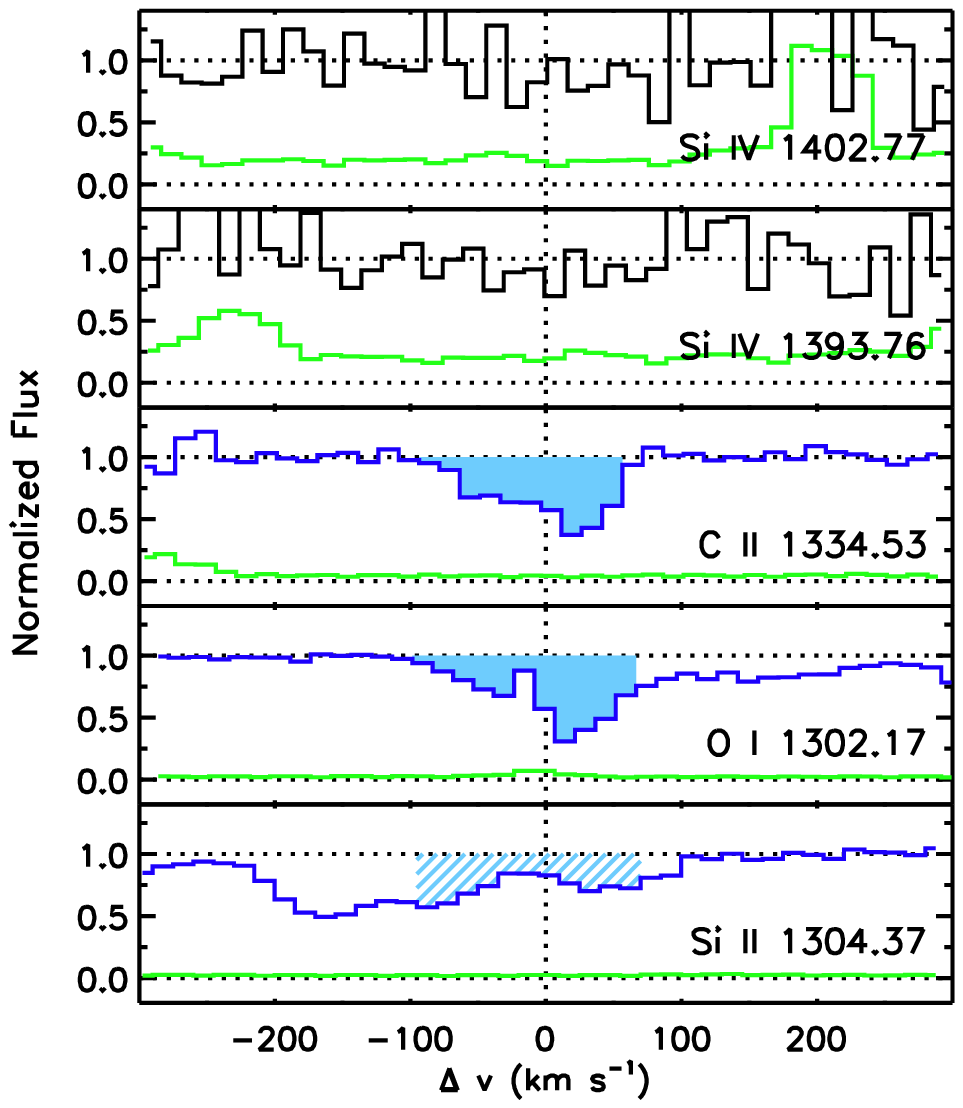}
   \caption{Stacked velocity plot for the absorption systems at $z=5.7529$ towards SDSS~J2315$-$0023.  The data are from ESI.  Lines are as in Figure~\ref{fig:0818_z5.7911}.  Note the change in velocity scale from Figures~\ref{fig:0818_z5.7911}-\ref{fig:1148_z6.2575}.  \siii~$\lambda1304$ is blended with an unrelated absorption line.
     \label{fig:2315_z5.7529}}
\end{figure}

At $z < 5$, low-ionization absorption systems are typically identified by their strong \hi\ absorption, particularly the extended damping wings of damped \lya\ systems \citep[DLAs; e.g.,][]{wolfe2005}.  At higher redshifts, however, absorption features in the \lya\ forest become sufficiently blended that identifying individual absorbers becomes difficult.  To identify low-ionization systems, we therefore searched redward of the forest for groups of metal lines including \siii~$\lambda 1260$, \siii~$\lambda 1304$, \oi~$\lambda 1302$, and \cii~$\lambda 1334$.  A detection required that two or more lines occur at the same redshift.  The lines were required to have similar velocity profiles, but the relative strengths of the different elements were not constrained.  For each sightline we searched between the QSO redshift and the redshift at which \oi\ enters the \lya\ forest.  The minimum and maximum redshifts covered are 5.33 and 6.42.  

In total we detected ten systems, five of which were reported in \citet{becker2006}.  Each contains \siii\ and \cii, and all but one contains detectable \oi.  Seven of the systems occur along sight lines for which we have high-resolution data.  The remainder fall towards QSOs we observed with ESI.  Five of the seven systems in the high resolution sample were reported in \citet{becker2006}, while the two additional high-resolution systems occur along a single sightline, SDSS~J0818$+$1722.  These two systems were also noted by \citet{dodorico2011a}.  The three systems in the ESI data occur along separate sight lines.

The systems are shown in Figures~\ref{fig:0818_z5.7911}-\ref{fig:2315_z5.7529}.  In addition to the low-ionization lines (\siii, \oi, and \cii), we plot the wavelength regions covering \siiv\ and \civ\ in cases where we have data.  \siiv\ may fall either in the optical or near-infrared, depending on the redshift of the system.  For \civ, our coverage comes entirely from our NIRSPEC data (Paper I).  The systems from \citet{becker2006} are reproduced primarily to demonstrate the apparent lack of high-ionization absorption lines, a point to which we will return later.  In the next sections we describe our basic measurements and completeness estimates.

\subsection{Equivalent Widths}\label{sec:ew}

\begin{deluxetable*}{lccccccc}
   \tabletypesize{} 
   \tablewidth{6.0in}
   \centering
   \tablecolumns{8}
   \tablecaption{Equivalent Widths} 
   \tablehead{\colhead{QSO} & 
              \colhead{$z_{\rm abs}$\tablenotemark{a}} &
              \colhead{\siii~$\lambda 1260$} &
              \colhead{\siii~$\lambda 1304$} &
              \colhead{\oi~$\lambda 1302$} &
              \colhead{\cii~$\lambda 1334$} &
              \colhead{\siiv~$\lambda 1394$\tablenotemark{b}} &
              \colhead{\civ~$\lambda 1548$\tablenotemark{c}} }
   \startdata
   SDSS~J1630$+$4012  &  5.8865  &  0.235    &  $< 0.076$  &  $< 0.031$  &  0.240  &  $< 0.362$  &  \nodata    \\ 
   SDSS~J2054$-$0005  &  5.9776  &  0.198    &     0.072   &     0.124   &  0.221  &  $< 0.096$  &  \nodata    \\ 
   SDSS~J2315$-$0023  &  5.7529  &  \nodata  &  $< 0.195$  &     0.238   &  0.240  &  $< 0.069$  &  \nodata    \\ 
   SDSS~J0818$+$1722  &  5.7911  &  0.267    &     0.056   &     0.182   &  0.230  &  $< 0.105$  &  $< 0.072$  \\
   SDSS~J0818$+$1722  &  5.8765  &  0.064    &  $< 0.008$  &     0.058   &  0.056  &  $< 0.044$  &  $< 0.038$  \\
   SDSS~J1623$+$3112  &  5.8415  &  \nodata  &     0.135   &     0.391   &  0.278  &  $< 0.060$  &  \nodata    \\ 
   SDSS~J1148$+$5251  &  6.0115  &  \nodata  &     0.038   &     0.162   &  0.154  &  $< 0.063$  &  $< 0.160$  \\
   SDSS~J1148$+$5251  &  6.1312  &  \nodata  &     0.016   &     0.079   &  0.067  &  $< 0.015$  &  $< 0.040$  \\
   SDSS~J1148$+$5251  &  6.1988  &  0.020    &  $< 0.028$  &     0.020   &  0.015  &  $< 0.072$  &  $< 0.113$  \\
   SDSS~J1148$+$5251  &  6.2575  &  0.045    &  $< 0.008$  &     0.036   &  0.047  &  $< 0.030$  &  \nodata
   \enddata 
   \tablecomments{Rest-frame equivalent widths are given in Angstroms.  Upper limits are 1$\sigma$.}
   \tablenotetext{a}{Minor differences from redshifts given in \citet{becker2006} are due to an improvement in the HIRES wavelength solution.}
   \tablenotetext{b}{This gives the smaller of the upper limit on \siiv~$\lambda
   1394$ and twice the upper limit on \siiv~$\lambda 1403$.}
   \tablenotetext{c}{This gives the smaller of the upper limit on \civ~$\lambda
   1548$ and twice the upper limit on \civ~$\lambda 1551$.}
   \label{tab:equiv_widths}
\end{deluxetable*}

We measured the rest-frame equivalent widths, $W_0$, for each available transition.  The results are given in Table~\ref{tab:equiv_widths}.  For detected low-ionization lines, the wavelength intervals over which the equivalent widths were measured are shown as shaded regions in Figures~\ref{fig:0818_z5.7911}-\ref{fig:2315_z5.7529}.  For non-detected low-ionization lines, upper limits are measured over the same intervals.  We define our upper limits to be the sum of any net absorption, as in the case of blends with unrelated lines, and the uncertainty in the normalized flux integrated over the chosen window.  

High-ionization lines are often kinematically distinct from low-ionization lines in lower-redshift DLAs and sub-DLAs \citep[e.g.,][]{fox2007a}.  We therefore place conservative upper limits on the equivalent widths of \siiv\ and \civ\ by integrating over a $\pm 100$~\kms\ window around the nominal system redshift.  We measure limits for both lines in the doublet separately, but take advantage of the fact that \ew\ for the stronger transition should be no more than twice that of the weaker transition.  For \siiv, the smaller of $W_{0}({\mbox \siiv}\,\lambda 1394)$ and $2W_{0}({\mbox \siiv}\,\lambda 1403)$ is given in Table~\ref{tab:equiv_widths}.  For \civ, we give the smaller of $W_{0}({\mbox \civ}\,\lambda 1548)$ and $2W_{0}({\mbox \civ}\,\lambda 1551)$.

\subsection{Column Densities}\label{sec:columns}

\begin{deluxetable*}{lcccccc}
   \tabletypesize{} 
   \tablewidth{6.0in}
   \centering
   \tablecolumns{7}
   \tablecaption{Column Densities for High-Resolution Systems} 
   \tablehead{\colhead{QSO} & 
              \colhead{$z_{\rm abs}$} &
              \colhead{$\log{N_{\msiii}}$} &
              \colhead{$\log{N_{\moi}}$} &
              \colhead{$\log{N_{\mcii}}$} &
              \colhead{$\log{N_{\msiiv}}$} &
              \colhead{$\log{N_{\mciv}}$} }
   \startdata
   SDSS~J0818$+$1722  &  5.7911  &  $13.41 \pm 0.03$  &  $ 14.54 \pm 0.03$  &  $ 14.19 \pm 0.03$  &  $<12.9$  &  $<13.4$  \\
   SDSS~J0818$+$1722  &  5.8765  &  $12.78 \pm 0.05$  &  $ 14.04 \pm 0.05$  &  $ 13.62 \pm 0.07$  &  $<12.6$  &  $<13.0$  \\
   SDSS~J1623$+$3112  &  5.8415  &  $14.09 \pm 0.02$  &  $>15.0$            &  $>14.3$            &  $<13.0$  &  \nodata  \\
   SDSS~J1148$+$5251  &  6.0115  &  $13.51 \pm 0.03$  &  $ 14.65 \pm 0.02$  &  $ 14.14 \pm 0.06$  &  $<12.7$  &  $<13.8$  \\
   SDSS~J1148$+$5251  &  6.1312  &  $13.29 \pm 0.09$  &  $ 14.79 \pm 0.34$  &  $ 13.88 \pm 0.22$  &  $<12.3$  &  $<13.2$  \\
   SDSS~J1148$+$5251  &  6.1988  &  $12.12 \pm 0.05$  &  $ 13.49 \pm 0.13$  &  $ 12.90 \pm 0.11$  &  $<12.9$  &  $<13.6$  \\
   SDSS~J1148$+$5251  &  6.2575  &  $12.67 \pm 0.03$  &  $ 13.97 \pm 0.15$  &  $ 13.56 \pm 0.14$  &  $<12.5$  &  \nodata  
   \enddata 
   \tablecomments{Column densities were computed using the apparent
   optical depth method in all cases except the $z=6.1312$ system
   towards SDSS~J1148$+$5251.  For this system, column densities were
   measured using Voigt profile fits, for which a Doppler parameter
   $b=5.7 \pm 1.0$~\kms\ was returned.}
   \label{tab:column_densities}
\end{deluxetable*}

Column densities of each ion were measured for systems in the high resolution sample, where the velocity structure is most likely to be resolved.  For six out of seven of these, column densities were measured using the apparent optical depth method \citep{savage1991}.  Here the optical depths of individual pixels, $\tau_{i}$, are measured, and the total column density in cm$^{-2}$ is given by 
\begin{equation}
   N = \frac{3.768 \times 10^{14}}{f\,\lambda_{0}} \sum_{i}{\tau_{i} \delta v_{i}} \, ,
\end{equation}
where $f$ is the oscillator strength, $\lambda_{0}$ is the rest wavelength in Angstroms, and $\delta v_{i}$ is the velocity width of a pixel in \kms.  This approach has the advantage that is does not require the absorption to be decomposed into individual components.  We integrated the optical depths over the shaded regions in Figures~\ref{fig:0818_z5.7911}-\ref{fig:1148_z6.2575}.   Prior to measuring the low-ionization lines, we smoothed our spectra by convolving the flux with a Gaussian kernel whose FWHM is equal to one half of the instrumental resolution.  This takes advantage of the fact that the spectra are known to have a finite resolution, which is roughly three pixels for our choice of binning.  Mild smoothing prevents individual noisy pixels with flux near or below  zero from dominating the summed optical depths, while only slightly degrading the spectral resolution (here by $\sim$12\%).  In cases where the spectra had high $S/N$, or for transitions which contained no pixels near zero, smoothing had a negligible effect on the measured column density.  We note, however, that smoothing should be used cautiously where the flux may genuinely go to zero.  For example, in the $z = 6.1312$ absorber towards SDSS~J1148$+$5251, \oi~$\lambda1302$ is marginally saturated.  We therefore fit Voigt profiles rather than apply the apparent optical depth method in this case.  This system appears to have a single narrow component, and so by requiring that each transition have the same redshift and Doppler parameter (i.e., assuming turbulent broadening), we can obtain a rough column density estimate for \oi\ even though the central two pixels show nearly zero flux.

Upper limits on the column densities were computed for non-detected species.  For low-ionization lines, the optical depths were integrated over the same velocity interval where other low-ionization species were detected.  Limits on \civ\ and \siiv\ were set by integrating the apparent optical depths over an interval -100\,\kms\ to +100\,\kms\ from the optical depth-weighted mean redshift of the low-ionization lines.  For the high-ionization doublets, the column density in the $i$th velocity bin was taken to be the inverse variance-weighted mean of the column densities measured from both transitions in cases where the measurements were consistent within the errors.  In cases where they were not consistent, as would be expected if one transition was blended with an unrelated line, the smaller of the two values was taken.

We used our optical depth measurements of \siii\ to infer another quantity used to classify low-ionization systems, the equivalent width of \siii~$\lambda1526$, \W\ \citep{prochaska2008a}.   Since we do not cover \siii~$\lambda1526$ with our current data,  the lines profile for this transition was reconstructed from the optical depths of \siii~$\lambda1260$ and/or $\lambda1304$.  In all cases where this was done, at least one of these transitions was covered and not saturated.  Since the optical depth should scale linearly with the oscillator strength, we therefore expect our estimates of \W\ to be reasonably accurate in all cases.   The results are given in Table~\ref{tab:vel_widths}.

\subsection{Velocity Widths}\label{sec:velocities}

\begin{deluxetable*}{lcccc}
   \tabletypesize{} 
   \tablewidth{4.0in}
   \centering
   \tablecolumns{5}
   \tablecaption{Line Widths for High-Resolution Systems} 
   \tablehead{\colhead{QSO} & 
              \colhead{$z_{\rm abs}$} &
              \colhead{$\Delta v_{90}$\tablenotemark{a}} &
	      \colhead{Transition\tablenotemark{b}} &
	      \colhead{$W_{1526}$\tablenotemark{c}} \\
	       & & (\kms) & & (\AA) }
   \startdata
   SDSS~J0818$+$1722  &  5.7911  &  134  &  \cii~$\lambda1334$   &  0.066  \\		       
   SDSS~J0818$+$1722  &  5.8765  &  25   &  \oi~$\lambda1302$    &  0.016  \\		       
   SDSS~J1623$+$3112  &  5.8415  &  223  &  \siii~$\lambda1304$  &  0.249  \\		       
   SDSS~J1148$+$5251  &  6.0115  &  41   &  \oi~$\lambda1302$    &  0.074  \\		       
   SDSS~J1148$+$5251  &  6.1312  &  17\tablenotemark{d}   &  \cii~$\lambda1334$  &  0.028  \\ 
   SDSS~J1148$+$5251  &  6.1988  &  17   &  \siii~$\lambda1260$  &  0.004  \\		       
   SDSS~J1148$+$5251  &  6.2575  &  13   &  \siii~$\lambda1260$  &  0.012                     
   \enddata 
   \tablenotetext{a}{Velocity width spanning 90\% of the optical depth in the low-ionization lines}
   \tablenotetext{b}{Transition used to measure $\Delta v_{90}$}
   \tablenotetext{c}{Rest equivalent width of \siii~$\lambda1526$ estimated
                     from optical depth measurements of available \siii\ lines}
   \tablenotetext{d}{Measured from Voigt profile fit}
   \label{tab:vel_widths}
\end{deluxetable*}

Finally, we measured the total velocity width of systems in our high resolution sample.  We use the velocity interval covering 90\% of the optical depth in low-ionization species, \dv, which has been used to characterize the kinematics of DLAs and sub-DLAs at lower redshifts \citep{prochaska1997}.  The widths are measured from the shaded pixels in Figures~\ref{fig:0818_z5.7911}-\ref{fig:1148_z6.2575}.   Mild smoothing was again applied.  The results are summarized in Table~\ref{tab:vel_widths}.  The transition used to measure \dv\ were chosen to have significant optical depth without being obviously saturated.  \oi\ was used only in cases where the velocity widths of \oi, \cii, and \siii\ are apparently equal.  In other cases, \cii\ and \siii\ appear to have absorption at high velocities not present in \oi, presumably due to a change in the ionization state of the gas.  

\subsection{Completeness}\label{sec:completeness}

\begin{figure}
   \epsscale{1.15} 
   \centering 
   \plotone{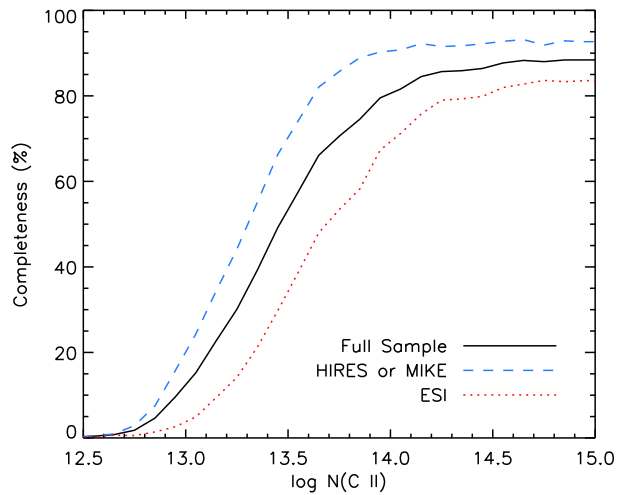}
   \caption{Completeness estimates as a function of \cii\ column density.  The solid line is the pathlength-weighted mean for the full the sample, while the dashed and dotted lines are for the high-resolution and ESI data, respectively.  See text for details.}
   \label{fig:completeness}
\end{figure}

We estimated our completeness by attempting to recover artificial absorption systems inserted into the data.  The synthetic lines were modeled as Voigt profiles with a Doppler parameter of 10\,\kms\ ($\Delta v_{90} = 23$\,\kms).  This choice focuses on the detectability of the narrower, weaker lines in our sample.  Artificial systems containing \oi, \cii, and \siii\ were placed at random into the data, assuming column density ratios equal to the mean values measured in the data (Section~\ref{sec:abundances}).  We then used an automatic routine to determine whether these systems would be detected.  The detection criteria were tuned to mimic the results of a by-eye identification.  A detection first required that a peak flux decrement greater than 1$\sigma$ be present at the same redshift for at least two of \siii~$\lambda1260$, \oi~$\lambda1302$, and \cii~$\lambda1334$ after the flux was smoothed by convolving with a Gaussian kernel of FWHM equal to the instrumental resolution.  For candidate pairs, the unsmoothed data were then fit using Gaussian profiles.  A detection required that the amplitude and FWHM for both lines were measured at greater than 3$\sigma$ confidence, and that the measured redshifts and velocity widths agreed with high confidence.  The ratio of the maximum optical depths in the two lines were also required to be within a factor of 0.4 to 2.0 of the ratio expected from the assumed relative abundances.

Each line of sight was tested with $10^{4}$ artificial systems over a range of column densities and covering the entire redshift interval used to search for low-ionization systems.  The pathlength-weighted mean results for the full sample are shown in Figure~\ref{fig:completeness}, along with separate results for the high-resolution and ESI data.  We estimate that we are 50\% complete at $\log{N(\mbox{\cii})} \simeq 13.3$ in the high-resolution data and $\log{N(\mbox{\cii})} \simeq 13.7$ in the ESI data.  Note that the completeness estimates do not go to 100\% for large column densities due to the fact that parts of the spectra are unusable due to strong atmospheric absorption or the presence of other absorption lines.  This is particularly an issue at redshifts where \siii~$\lambda1260$ is in the \lya\ forest, and the system must be identified from \oi~$\lambda1302$ and \cii~$\lambda1334$ alone.  Realistically, however, strong systems tend to have multiple components, making them easier to detect, and often have significant \siii~$\lambda1304$ absorption.

\section{Sample Properties}\label{sec:sample}

\subsection{Number Density}\label{sec:num_density}

\begin{figure*}
   \epsscale{0.85} 
   \centering 
   \plotone{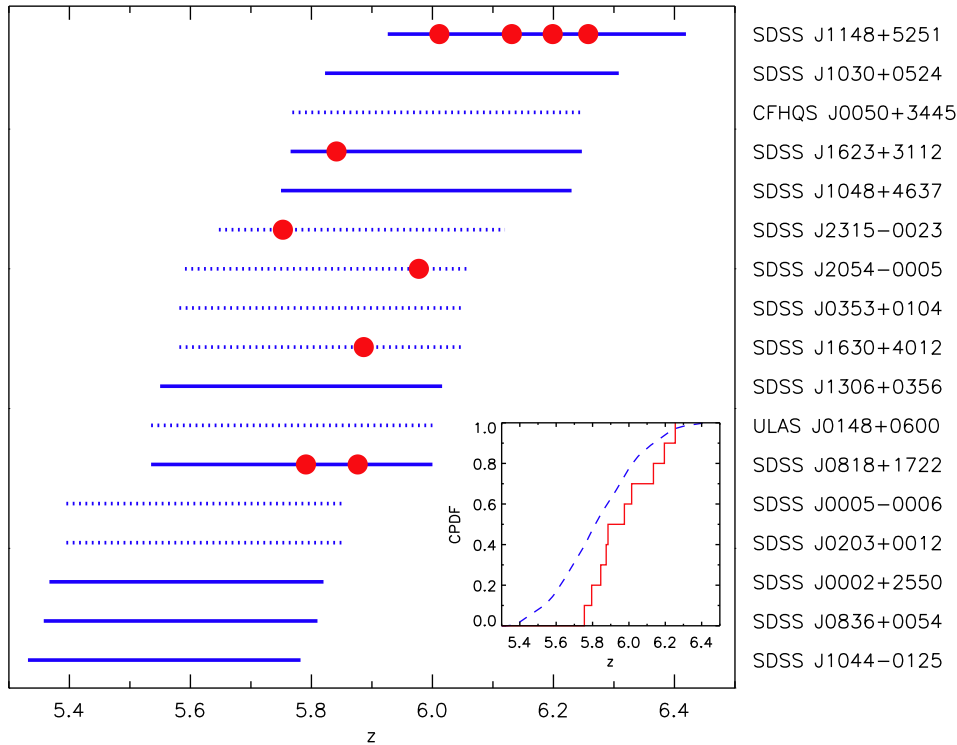}
   \caption{A schematic representation of our survey results.  The horizontal lines show the redshift interval surveyed for each QSO.  Solid lines are for HIRES or MIKE data.  Dotted lines are for ESI data.  Filled circles represent detected low-ionization systems.  The inset shows the cumulative probability density function (CPDF) for the survey pathlength, $\Delta X(z)$ (dashed line), and for the redshifts of the low-ionization systems (histogram).}
   \label{fig:oi_sightlines}
\end{figure*}

We now address the number density of low-ionization absorption systems at $z \sim 6$.  We compute the number density of absorbers per unit absorption pathlength interval, $\ell(X) \equiv d\mathcal{N}/dX$.  Here, $X$ is defined as
\begin{equation}
   X(z) = \int_{0}^{z} (1+z')^2 \frac{H_{0}}{H(z')} dz' \, ,
   \label{eq:X}
\end{equation}
where $H(z)$ is the Hubble constant at a given redshift \citep{bahcall1969}.  A non-evolving population of absorbers will have a constant number density per unit $X$.  Our total pathlength is $\Delta X = 39.5$ ($\Delta z = 8.0$, or $\Delta l = 3.6$~comoving Gpc).  The high-resolution data cover $\Delta X = 21.0$, while the ESI data cover $\Delta X = 18.5$.  Uncorrected for completeness, the line of sight number density in the entire sample is $\ell(X) = 0.25^{+0.21}_{-0.13}$  (95\% confidence), where the error bars include Poisson uncertainties only.  

Corrections to $\ell(X)$ for completeness are small compared to the errors, except when considering our weakest system.  The $z = 6.1988$ system towards SDSS~J1148$+$5251, which has $\log{N({\mbox \cii})} = 12.9$, would have been detected along only 9\% of our pathlength.  Few systems that are similar are known at lower redshifts.  The only comparable system of which we are aware is the sub-DLA at $z = 3.7$ found by \citet{peroux2007}, which has $\log{N({\mbox \cii})} = 13.1$ yet still contains \oi.  It is intriguing to consider whether more weak systems may be found at $z \sim 6$ in future, more sensitive surveys.

Seven out of our ten systems are detected in the high-resolution data, which suggests that the two lines of sight with multiple systems may be unusual.  At a given signal-to-noise ratio, however, HIRES and MIKE are more sensitive to narrow absorption systems than ESI, which has considerably lower resolution.  In order to determine whether the two data sets are consistent, we first take the ratio of the completeness estimates in Figure~\ref{fig:completeness} at the column densities of the seven systems in our high resolution sample.  Significantly, the weakest system ($\log{N({\mbox \cii})} = 12.90$) would be nearly undetectable in the ESI data, while the completeness of the ESI data for the two systems with $\log{N({\mbox \cii})} \simeq 13.6$ would be roughly half that of the high-resolution data.  If we also take into account the slightly shorter pathlength of the ESI data, and the fact that the ESI data are somewhat less sensitive to even the stronger systems, then we would expect to detect $\sim$4 systems in the ESI data if the high-resolution systems represent a fair sample.  The fact that we detected three systems with ESI suggests that the two data sets are broadly consistent.

It is not entirely straightforward to compare the number density of low-ionization absorption systems at $z \sim 6$ to lower redshifts.  At $z < 5$, low-ionization systems are generally identified by their strong \hi\ absorption (e.g., DLAs).  A systematic search for low-ionization lines independent of \hi\ column density, as we are forced to do here, has not been performed at lower redshifts.  We can estimate the number density of lower-redshift low-ionization systems, however, by identifying the \hi\ column density where low-ionization lines, particularly \oi, start to appear.  \oi\ is ubiquitous among DLAs \citep[$\log{N({\mbox \hi})} \ge 20.3$; e.g.,][]{wolfe2005}, and is also present in sub-DLAs down to $\log{N({\mbox \hi})} \simeq 19.0$ \citep{d-z2003,peroux2007}.  Low-ionization lines are also seen in lower column density systems.  However, these tend to show very strong \siiv\ and \civ\ absorption relative to \siii\ and \cii\ \citep[e.g.,][]{prochaska1999,prochter2010}, which is not seen in our systems (except, perhaps, for the $z=5.8865$ system towards SDSS~J1630$+$4012, which may contain strong \siiv).  We therefore conclude that our $z \sim 6$ low-ionization systems are most naturally compared to lower-redshift DLAs and sub-DLAs with $\log{N({\mbox \hi})} \ge 19.0$.  A more objective comparison must await either a blind survey for low-ionization metal lines at lower redshifts, or a more complete characterization of the metal line properties of sub-DLAs and lower column density systems.  In the following sections we generally compare our $z \sim 6$ systems to lower-redshift sub-DLAs, since they are more common than DLAs and therefore more likely to make up the majority of our high-redshift sample.

The number density of DLAs has been well established using the large pathlength available from the Sloan Digital Sky Survey \citep{prochaska2009,noterdaeme2009}.  Surveys using higher-resolution spectra have meanwhile been used to determine the number density of sub-DLAs \citep{peroux2005,omeara2007,guimaraes2009}.  The net result is that the total number of systems with $\log{N({\mbox \hi})} \ge 19.0$ is $\ell(X) \simeq 0.3$ over $3 < z < 5$.  This is very similar to the number density of low-ionization systems we find over $5.3 < z < 6.4$, which suggests that the number density of low-ionization systems may be roughly constant over $3 < z < 6$.  This contrasts to the total number density of optically thick Lyman limits systems ($\log{N({\mbox \hi})} \ge 17.2$), whose number density per unit $X$ increases by a factor of $\sim$2 from $z \sim 3$ to 6 \citep{songaila2010}.  It also contrasts to the number density of highly ionized metal absorption systems seen in \civ, which declines by at least a factor of four from $z \sim 3$ to 6 (Paper I).

\subsubsection{Possible Redshift Evolution}

A schematic representation of the redshifts of the detected systems is shown in Figure~\ref{fig:oi_sightlines}.  We also plot the cumulative probability density function of the survey pathlength, $\Delta X(z)$ (not adjusted for completeness), and for the redshifts of the low-ionization systems.  One peculiar feature is that all ten of our systems lie at $z \ge 5.75$, even though nearly 40\% of our pathlength is at lower redshifts.  We performed a K-S test to determine whether the distribution of absorption redshifts is consistent with our pathlength-weighted survey redshift distribution function.  The test returns a $D$ statistic of 0.39, which would be larger in a random sample of ten redshifts drawn from our survey path only 7\% of the time.  The distribution appears more unusual if we consider that, for a uniform distribution, the probability that all ten systems would randomly occur at $z > 5.75$ is only 0.7\%.  This ignores clustering, however.  If we count only one of the systems towards both SDSS~J1148$+$5251 and SDSS~J0818$+$1722 (i.e., seven `groups' of systems, the maximum effect of clustering), then the probability increases to 3\%.  We note that completeness should not be a significant factor, as our completeness is somewhat better below $z = 5.75$ than above.  Thus, there is evidence that the number density of low-ionization systems may be increasing with redshift over $z \sim 5$ to 6.  An expanded search at $z < 5.7$ would help to clarify this.

\subsection{Mass Density}\label{sec:mass_density}

\begin{deluxetable}{lcc}
   \tabletypesize{} 
   \tablewidth{2.0in}
   \centering
   \tablecolumns{5}
   \tablecaption{Mass Densities over $5.3 < z < 6.4$} 
   \tablehead{\colhead{Ion} & 
              \colhead{$\Omega_{\rm ion}$} &
              \colhead{$[{\rm M/H}]_{\rm global}$}  }
   \startdata
   \siii  &  $\ge 0.4 \times 10^{-8}$  &  $\ge -3.9$  \\
   \oi    &  $\ge 3.6 \times 10^{-8}$  &  $\ge -3.8$  \\
   \cii   &  $\ge 0.9 \times 10^{-8}$  &  $\ge -4.1$
   \enddata 
   \label{tab:omega}
\end{deluxetable}

\begin{deluxetable*}{lccccc}
   \tabletypesize{} 
   \tablewidth{4.0in}
   \centering
   \tablecolumns{5}
   \tablecaption{Relative Abundances for High-Resolution Systems} 
   \tablehead{\colhead{QSO} & 
              \colhead{$z_{\rm abs}$} &
	      \colhead{$(v_{\rm lo},v_{\rm hi})$\tablenotemark{a}} &
	      \colhead{[Si/O]} &
              \colhead{[C/O]} \\
               & & (\kms) & & }
   \startdata
   SDSS~J0818$+$1722  &  5.7911  &  (-86,45)  &  $-0.03 \pm 0.05$  &  $-0.14 \pm 0.04$  \\
   SDSS~J0818$+$1722  &  5.8765  &  (-16,12)  &  $-0.11 \pm 0.07$  &  $-0.16 \pm 0.09$  \\
   SDSS~J1623$+$3112  &  5.8415  &  (0,104)   &  $~~0.27 \pm 1.00$  &  $-0.30 \pm 1.41$  \\
   SDSS~J1148$+$5251  &  6.0115  &  (-36,32)  &  $~~0.01 \pm 0.04$  &  $-0.24 \pm 0.06$  \\
   SDSS~J1148$+$5251  &  6.1312  &  (-12,15)  &  $-0.35 \pm 0.35$  &  $-0.64 \pm 0.40$  \\
   SDSS~J1148$+$5251  &  6.1988  &  (-12,10)  &  $-0.23 \pm 0.14$  &  $-0.33 \pm 0.17$  \\
   SDSS~J1148$+$5251  &  6.2575  &  (-10,11)  &  $-0.15 \pm 0.15$  &  $-0.14 \pm 0.20$  \\
   \hline
   Mean\tablenotemark{b}  &      &            &  $-0.03 \pm 0.03$  &  $-0.18 \pm 0.03$
   \enddata 
   \label{tab:rel_abundances}
   \tablenotetext{a}{Minimum and maximum velocities with respect to the absorber redshift
                     over which column densities were integrated}
   \tablenotetext{b}{Inverse variance-weighted mean values excluding the $z=5.8415$ system,
                     for which \oi\ and \siii~$\lambda1304$ appear to be blended with
                     a lower-redshift \civ\ absorber}
\end{deluxetable*}

The comoving mass density of various ions can be characterized as a fraction of the critical density at $z = 0$, $\rho_{\rm crit}$, as
\begin{equation}
   \label{eq:omega}
   \Omega_{\rm ion} = \frac{H_0\,m_{\rm ion}}{c\,\rho_{\rm crit}}
                 \int{N_{\rm ion} \, f(N_{\rm ion},X) \, dN_{\rm ion}} \, ,
\end{equation}
where $f(N_{\rm ion},X) \equiv \partial ^2{\mathcal N}/\partial N_{\rm ion}\partial X$.  The integral in Eq.~\ref{eq:omega} is often approximated using the observed column densities within a sample as 
\begin{equation}
   \int{N_{\rm ion} \, f(N_{\rm ion},X) \, dN_{\rm ion}} \simeq \frac{\sum{N^{\rm obs}_{\rm ion}}}{\Delta X} \, .
   \label{eq:approx}
\end{equation}
The validity of this approximation, however, is highly model-dependent.  A power law $f(N_{\rm ion},X)$, for example, may hide most of the mass in rare, high column density systems.  It is unclear what shape the distribution of low ion column densities should have, as it will depend on the underlying hydrogen column density distribution, the distribution of metallicities, and ionization corrections that are potentially important for sub-DLAs or somewhat weaker systems.  The distribution may be log-normal, as there are several multiplicative factors at work.  The present sample is too small, however, to characterize the distribution with any confidence.  Given that much of the metal mass in the low-ionization phase may reside in rare systems that have both high \hi\ column densities and higher than average metallicities, a conservative approach is to use the present sample to set lower limits on $\Omega_{\rm ion}$ using the approximation in Eq~\ref{eq:approx}.  The values for \siii, \oi, and \cii\ are given in Table~\ref{tab:omega}.  We also convert these values into minimum mass-averaged global metallicities by dividing the ionic densities by the total mass density of hydrogen at $z = 6$, assuming \citet{asplund2005} solar abundances.  Notably, $\Omega({\mbox \cii})$ is twice as high as $\Omega({\mbox \civ})$ measured by \citet{rw2009}, although both were derived from small samples of lines.  The true values of $\Omega_{\rm ion}$ may be factors of several higher in both cases.

\subsection{Relative Abundances}\label{sec:abundances}

At $z \sim 6$ the \lya\ forest is sufficiently absorbed that \hi\ column densities cannot be measured for individual absorbers, so we are unable to directly measure metallicities for these systems.  We can, however, determine relative abundances using only the metal lines.  The relative abundances for the systems in our high resolution sample are given in Table~\ref{tab:rel_abundances}.  These are measured with respect to the solar values from \citet{asplund2005}, and assume no dust depletion and no ionization corrections such that, for example, $[{\rm Si/O}] = {\rm log}[N({\mbox \siii})/N({\mbox \oi})] - {\rm log}[n({\rm Si})/n({\rm O})]_\odot$.  In order to minimize any potential ionization corrections, we use column densities measured only over the velocity interval that contains \oi.  In the case of the $z=5.8415$ system towards SDSS~J1623$+$3112, the \oi\ and \siii~$\lambda1304$ lines appear to be blended with a \civ\ system at $z=4.754$.  The inverse variance-weighted mean values for the remaining six systems are $[{\rm Si/O}] = -0.03 \pm 0.03$ and $[{\rm C/O}] = -0.18 \pm 0.03$ (1$\sigma$).  We note that the scatter in the individual values is entirely consistent with the measurement errors, which suggests that the ionization corrections may indeed be small.  The mean values are similar to the values reported in \citet{becker2006}, as expected since many of the systems in Table~\ref{tab:rel_abundances} were included in the earlier sample.  As noted there, the relative abundances are consistent with enrichment from Type II supernovae of low-metallicity massive stars \citep{woosley1995,chieffi2004}. Our [C/O] values are also consistent with those found by \citet{pettini2008} for metal-poor DLAs at $z \sim 2-3$ (but see \citet{penprase2010}, who find higher values), and with the [C/O] values of low-metallicity stars in the Galactic halo \citep{akerman2004}.  We will further address the implications of the relative abundances for stellar populations at high redshift in a forthcoming paper.

\subsection{High-Ionization Lines}\label{sec:high_ion}

\begin{figure*}
   \epsscale{1.0} 
   \centering 
   \plotone{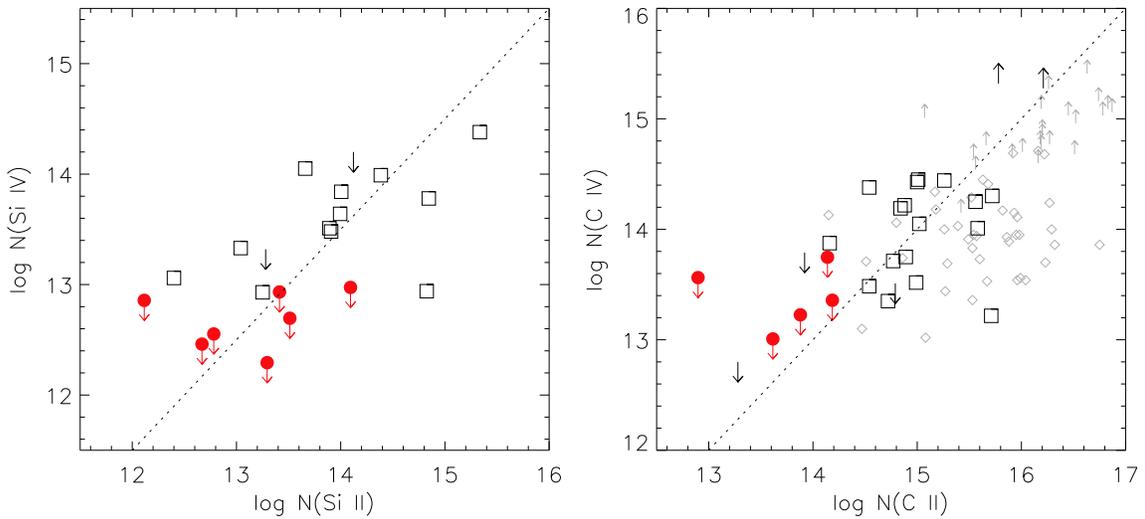}
   \caption{High-ionization (\siiv\ or \civ) column densities as a function of low-ionization (\siii\ or \cii) column densities.  Filled circles with arrows are upper limits for the $z \sim 6$ systems in our high resolution sample for which \siiv\ or \civ\ coverage is available.  Other symbols are for $z \sim 2-4$ sub-DLAs (squares and large arrows) and DLAs(diamonds and small arrows) in the literature.  The \siii\ and \cii\ column densities for the sub-DLAs and DLAs are either published measurements or were estimated from published metallicities and $N({\mbox \hi})$ values.  See text for details.  A linear relation is plotted as a dotted line in each panel.  For silicon, the relation is $\log{N({\mbox \siiv})} = \log{N({\mbox \siii})} - 0.5$.  For carbon, the relation is $\log{N({\mbox \civ})} = \log{N({\mbox \cii})} - 1.0$.  The relations are not fits, but are meant to demonstrate the direction of equal fractional changes in the metal column densities of the low- and high-ionization phases.  In this sense, the $z \sim 6$ systems are consistent with being lower-metallicity analogues of the $z < 4$ sub-DLAs and DLAs.  We note that sub-DLAs are likely to be more common than DLAs at $z \sim 6$, just as they are at lower redshifts, and so comparing our systems with lower-redshift sub-DLAs may be the most appropriate choice.}
   \label{fig:highion}
\end{figure*}

A notable feature of our low-ionization systems is the lack of strong absorption by high-ionization species (\siiv\ and \civ).  At $z \sim 6$, \siiv\ and \civ\ are shifted into the far red and near infrared, and we stress that our spectra at these wavelengths are generally of modest quality.  In most cases, however, the high-ionization lines must be significantly weaker than the low-ionization lines.

Significant \civ\ absorption is seen for nearly all lower-redshift DLAs and sub-DLAs.  It is kinematically distinct from the low-ionization lines \citep{wolfe2000a}, and has been interpreted as evidence for galactic winds \citep{fox2007a}.  We therefore investigate whether the non-detections of \siiv\ and \civ\  infer that the host galaxies of our $z \sim 6$ systems lack strong winds or other nearby, highly-ionized gas.  The column densities of high-ionization lines are plotted verses the column densities of low-ionization lines in Figure~\ref{fig:highion} for our $z \sim 6$ systems and a sample of DLAs and sub-DLAs at $z < 4.2$ drawn from the literature \citep{d-z2003,peroux2007,fox2007a}.  For the literature data, direct measurements of $N({\mbox \siii})$ are used when available.  Otherwise, $N({\mbox \siii})$  was estimated from the \hi\ column density and metallicity measurements, assuming solar abundances and that all of the silicon in the low-ionization phase is in \siii\ (i.e., no ionization corrections).  For the 19 sub-DLAs with \siii\ measurements, which had $19.1 < \log{N({\mbox \hi})} < 20.2$, this procedure reproduced the published $N({\mbox \siii})$ to within 0.03 dex in all cases.  This is not surprising, since the metallicity estimates were often based on \siii.  Column densities were estimated for \cii\ using the same procedure.  For one sub-DLA, a limit on \civ\ that was found to be unusually low was re-estimated from the published spectrum.  Otherwise, the \siiv\ and \civ\ measurements are as originally reported.

The systems at $z \sim 6$ tend to show lower high-ionization column densities than DLAs and sub-DLAs at lower redshifts.  However, the upper limits on \siiv\ and \civ\ are consistent with a trend of declining high-ionization line strength with declining low-ionization column density.  This is reminiscent of the correlation between \civ\ strength and metallicity identified by \citet{fox2007a} for DLAs and sub-DLAs at $z < 3.6$.  The trend suggests that the $z \sim 6$ systems may lack strong high-ionization lines because they have low metallicities.  Highly-ionized envelopes or winds may still be present but not detected in absorption with present-quality data.

\citet{rw2009} and \citet{dodorico2011a} noted a possible \civ\ doublet at $z \simeq 5.79$ towards SDSS~J0818+1722, which would coincide with one of our low-ionization systems.  There is possible \civ\ absorption in our NIRSPEC data near $\Delta v = -100$~\kms, where there is also \siii\ and \cii\ absorption but no \oi.  The lack of \oi\ suggests that the gas at this velocity may be ionized, and so some \civ\ might be expected.  Confirmation must await higher quality data.

\subsection{Velocity Width Distribution} \label{sec:vel_distrib}

\begin{figure}
   \epsscale{1.15} 
   \centering 
   \plotone{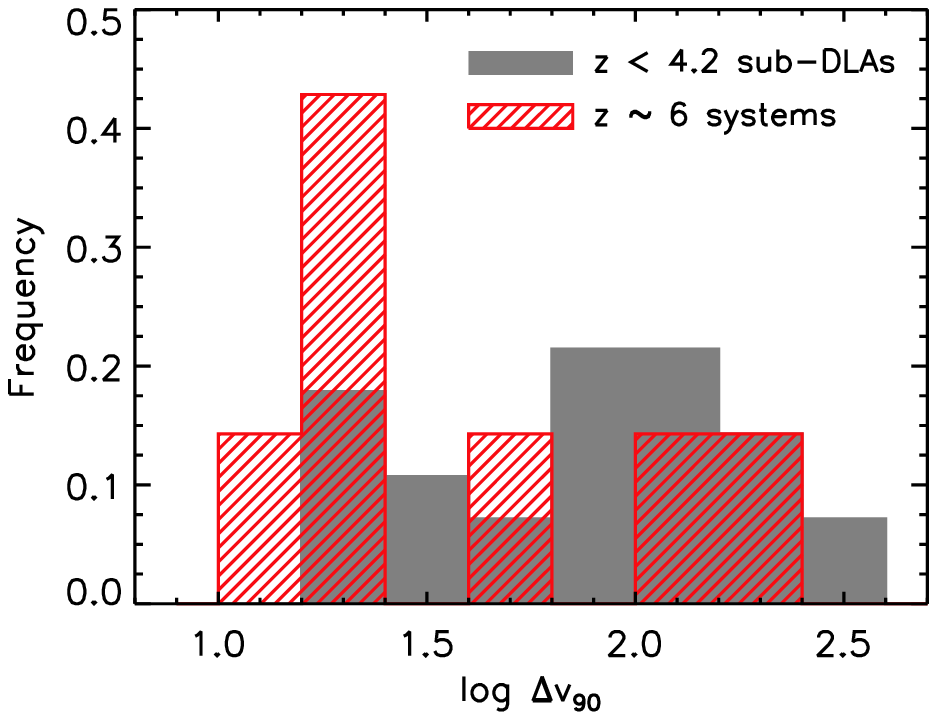}
   \caption{Distribution of \dv\ values for the seven $z \sim 6$ systems with high-resolution data (shaded histogram) and for lower-redshift sub-DLAs (filled histogram).  The 28 sub-DLA measurements were computed from data in \citet{d-z2003} and \citet{peroux2007}.  There is some indication that narrow systems are more common at $z \sim 6$.  Both samples are small, however, and the overall distributions are statistically not significantly different.}
   \label{fig:dv}
\end{figure}

The absorbers in our sample show a range of velocity profiles, from single, narrow lines to broad, multi-component systems.  The distribution of \dv\ values among the seven systems in our high resolution sample is plotted in Figure~\ref{fig:dv}.  We also show the distribution of \dv\ for 28 sub-DLAs in the literature \citep{d-z2003,peroux2007,fox2007a}.  For the systems in \citet{d-z2003} and \citet{peroux2007}, the velocity widths were calculated from published Voigt profile fits.  There is some indication that the low-ionization systems at $z \sim 6$ are more likely to be narrow.  A K-S test, however, indicates that there is a reasonable likelihood (27\%) that the two samples were drawn from the same parent distribution.

\subsection{Mass-Metallicity Relation}\label{sec:metallicities}

\begin{figure*}
   \epsscale{1.1} 
   \centering 
   \plotone{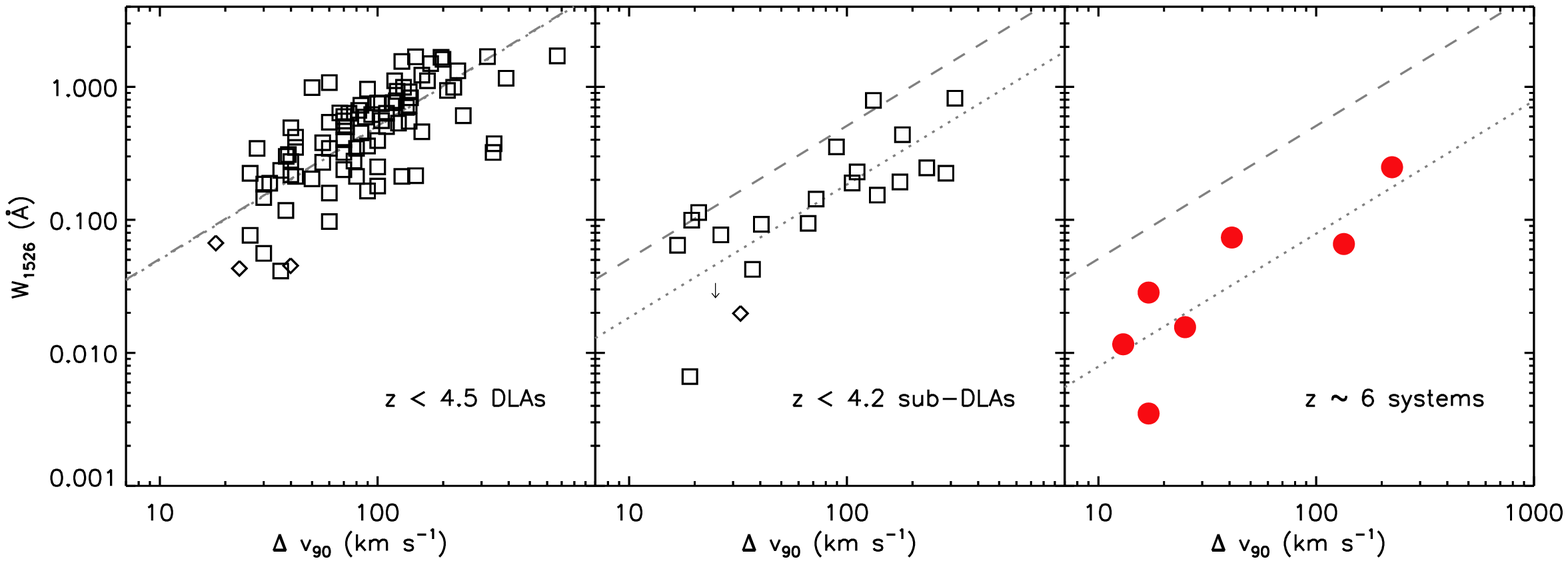}
   \caption{Rest equivalent width of \siii~$\lambda1526$ as a function of \dv.  The left and middle panels are for $z < 4.5$ DLAs and $z < 4.2$ sub-DLAs, respectively, from the literature.  These include the metal-poor systems from \citet{pettini2008}, which are shown as diamonds. See text for details.  The right panel is for our $z \sim 6$ systems with high-resolution data.  The dashed line in each panel shows the relation expected for fully saturated absorption, $W^{\rm sat}_{1526} = (1526.7\,{\mbox \AA})\Delta v_{90}/c$.  The dotted lines show this relation offset vertically such that the mean deviation in $\log{W_{1526}}$ is zero.  For the DLAs the offset is close to zero.  At a given velocity width, the metal lines for the $z \sim 6$ systems are significantly weaker than for either the lower-redshift DLAs or sub-DLAs.}
   \label{fig:w1526_dv}
\end{figure*}

The high opacity of the \lya\ forest at $z \sim 6$ prevents us from measuring \hi\ column densities for these systems.  We are therefore unable to obtain metallicities directly.  Previous works have noted correlations between metallicity and either the velocity width \citep{ledoux2006} or the rest equivalent width of \siii~$\lambda 1526$ \citep[\W;][]{prochaska2008a}.  These relationships have only been calibrated for DLAs, however, while our systems potentially have lower \hi\ column densities.  Some insight can be gained, however, by comparing the relationship between \dv\ and \W\ in our study to systems at lower redshifts.

Figure~\ref{fig:w1526_dv} shows the relationship between \dv\ and \W\ for DLAs and sub-DLAs at $z < 4.5$ and for our low-ionization systems at $z \sim 6$.  The DLA data are from \citet{prochaska2008a}.  Values of \dv\ and \W\ for sub-DLAs were computed from published Voigt profile fits from \citet{d-z2003} and \citet{peroux2007}.  We also include the three metal-poor DLAs and one metal-poor sub-DLA from \citet{pettini2008}.  The $z \sim 6$ sample includes the seven systems in our high resolution sample.

Following \citet{prochaska2008a}, we plot the expected relation for fully saturated absorption, $W^{\rm sat}_{1526} = (1526.7\,{\mbox \AA})\Delta v_{90}/c$, as a dashed line in each panel.  We also show this relation scaled by a factor such that mean residual in $\log{W_{1526}}$ is zero for each data set.  The DLAs roughly follow the relation expected for saturated absorption, while \W\ for the sub-DLAs is lower by 0.44 dex.  For the $z \sim 6$ systems, the offset is 0.81 dex, or a factor of $\sim$2 lower than the sub-DLAs.  At a given velocity width, therefore, the low-ionization systems have significantly weaker metal lines than either DLAs or sub-DLAs at $z < 4.5$.

We can translate the offset in \W\ into an approximate difference in metallicity.  \citet{prochaska2008a} fit a relation for DLAs of the form $[{\rm M/H}] = a + b\log{W_{1526}}$, finding $(a,b)_{\rm DLA} = (-0.92, 1.41)$ with an $r.m.s.$ scatter of 0.25 dex.  We have fit a similar relation for the sub-DLAs using the published metallicities, finding $(a,b)_{\rm sub{\mbox-}DLA} = (-0.28 \pm 0.15, 1.19 \pm 0.18)$, with a scatter of 0.33 dex.  It is apparent, therefore, that the metallicity estimate will depend on the \hi\ column density of the absorber.  If we assume that our $z \sim 6$ systems are at least sub-DLAs, then the decline in \W\ suggests that the metallicity of the $z \sim 6$ is systems at a given \dv\ is lower by $\gtrsim$0.4 dex than that of lower-redshift systems. 

The velocity width is expected to correlate with the mass of the dark matter halo \citep[e.g.,][]{haehnelt1998,maller2001,prochaska2008a,pontzen2008}.  This decline in \W\ at a given \dv\ can therefore be interpreted as evolution in the mass-metallicity relation for the host galaxies of low-ionization absorbers.   \citet{ledoux2006} noted a similar redshift evolution for DLAs in their [M/H]-\dv\ relation, in which the metallicity at a given \dv\ declined by $\sim$0.3 dex from $1 .7 < z < 2.43$ to $2.43 < z < 4.3$.  Our results demonstrate that this decline in the mass-metallicity relation likely extends to galaxies near the reionization epoch.

\section{Discussion}\label{sec:discussion}

We now consider whether the number density of low-ionization systems is expected to remain roughly constant with redshift, as observed.  Previous works have shown that the number density (per unit X) of both DLAs and sub-DLAs shows little evolution over $3 \le z \le 5$ \citep{peroux2005,prochaska2005,prochaska2009,noterdaeme2009,guimaraes2009}.  Our observations suggests that the number density remains roughly constant out to $z \sim 6$, but why should this happen?  The number density of a population of absorbers will depend on their comoving number density, $\phi_{\rm c}$, and their mean physical absorption cross section, $\langle \sigma_{\rm abs} \rangle$, as
\begin{equation}
   \ell(X) = \frac{c}{H_0} \, \phi_{\rm c} \, \langle \sigma_{\rm abs} \rangle \, .
   \label{eq:number_density}
\end{equation}
Observational \citep[e.g.,][]{cooke2006a,cooke2006b} and theoretical \citep[e.g.,][]{gardner1997a,haehnelt1998,nagamine2004a} arguments generally suggest that DLAs are hosted by dark matter haloes of mass $M \gtrsim 10^{8-9}$~\Msun, and may be largely associated with halos in the range $10^9 < M/M_{\odot} < 10^{11}$ \citep{pontzen2008}.  For a Sheth-Tormen halo mass function, $\phi_{\rm c}$ for $10^{10}$~\Msun\ halos, for example, will be lower at $z=6$ than at $z=3$ by a factor of $\sim$4 \citep{sheth2001,reed2007}.  It is possible that the host halos of low-ionization systems are somewhat less massive at higher redshifts, which would increase their space density.  The similarity of the velocity width distributions of the $z \sim 6$ systems and low-redshift sub-DLAs suggests that the halo masses are similar, although the velocity width may only be moderately sensitive to the mass.  For an NFW dark matter profile \citep{navarro1997}, the virial velocity scales with the halo mass as $v_{\rm vir} \propto M_{\rm vir}^{1/3}H(z)^{1/3}$, where $H(z)$ is the Hubble parameter at redshift $z$.  If \dv\ scales with the virial velocity \citep[e.g.,][]{haehnelt1998}, then a decrease in the typical halo mass by a factor of ten from $z=3$ to 6 would only produce a decline in the typical \dv\ of $\sim$0.2~dex.  Such a decline is consistent with the present data, but as noted in Section~\ref{sec:vel_distrib}, the distribution of \dv\ values for the $z \sim 6$ sample is formally consistent with that of lower-redshift sub-DLAs.  A larger sample is needed to determine whether there is a genuine decrease in the mean velocity width.

Multiple factors may increase the absorption cross-section with redshift.  At a given mass, an NFW profile will have a higher density, particularly in the inner regions.  The external UV background also decreases from $z \sim 3$ to 6 \citep{fan2006b,bolton2007b,wyithe2010,calverley2010}.  Both of these factors should increase the radius out to which a halo of a given mass can become self-shielded.  

We explored simple models using NFW profiles to predict how the cross-section of low-ionization gas should evolve with redshift.  Assuming that the gas traces the dark matter and is in photoionization equilibrium, it is straightforward to predict the radius at which the gas becomes self-shielded and to estimate the cross-section over which the projected \hi\ column density exceeds a given value.  For this we assumed a gas temperature of 20,000~K and a metagalactic \hi\ ionization rate $\Gamma = 10^{-11.9}~{\rm s^{-1}}$ ($10^{-12.8}~{\rm s^{-1}}$) at $z = 3$ (6).  This simple model suggests that the cross section over which a halo of a given mass would be a DLA should scale roughly linearly with the halo mass, which agrees broadly with recent theoretical models \citep{nagamine2004a,pontzen2008,tescari2009}.  It also predicts that the DLA cross section for a given mass halo should increase by a factor of $\sim$4 from $z = 3$ to 6.  If DLAs are mainly hosted by dark matter halos with masses $M \ge 10^9$~M$_\sun$, then this increase in the cross section will offset the decline in the comoving spatial density to produce a roughly constant number density of DLAs with redshift.  The same is true for sub-DLAs in this model.  We stress that this is an extremely simple model, and it is unlikely to reproduce detailed observations such as the column density or velocity width  distributions of DLAs.  Nevertheless, it gives some assurance that the apparent lack of evolution in the number density of low-ionization absorbers can be plausibly explained by a decline in the spatial density of dark matter halos combined with an increase in the mean absorption cross section.  More detailed analysis must be left for future work.
  
The high number density of low-ionization systems makes it likely that they arise from galaxies below the detection limits of current galaxy surveys.  Integrating the $z \sim 6$ luminosity function of $i$-dropout galaxies from \citet{bouwens2007} down to the limit of the Hubble Ultra Deep Field ($\sim$0.04$L^{*}_{z=3}$) gives a space density $\phi(M_{\rm AB} \le -18) = 5 \times 10^{-3}$ per comoving Mpc$^3$.  For $\ell(X) \simeq 0.25$, this implies a mean absorption cross-section $\langle \sigma_{\rm abs} \rangle \simeq 1 \times 10^4~{\rm kpc}^{-2}$, or $\langle R_{\rm abs}\rangle \simeq 60$~kpc for a circular projected geometry.  The \lya-emitting galaxies (LAEs) at $z \sim 6$ included in current surveys would need an even larger cross-section.  \citet{ouchi2008} found a density of LAEs at $z=5.7$ of $\sim$$7 \times 10^{-4}$ per Mpc$^{-3}$ down to a \lya\ luminosity of $2.5 \times 10^{42}~{\rm erg~s^{-1}}$.  These LAEs would require $\langle R_{\rm abs}\rangle \simeq 160$~kpc to account for the observed number density of low-ionization absorption systems.  In contrast, constraints from paired QSO lines of sight suggest a typical DLA size at $z = 1$-2 of $R_{\rm DLA} \simeq 5-10$~kpc  \citep{monier2009,cooke2010}, although in some cases they may be larger \citep[e.g.,][]{briggs1989}.  The sizes of sub-DLAs are less well constrained, but they may plausibly be a factor of two larger than DLAs (see, for example, the scaling relation in \citet{monier2009}, or Figure~2 in \citet{pontzen2008}).  In that case, they would still need to be significantly larger at $z = 6$ to arise solely from galaxies similar to those in current surveys.  It seems more plausible, therefore, that these low-ionization systems trace fainter, more numerous sources.  This would be consistent with the results of \citet{rauch2008}, who find a likely connection between DLAs and a population of ultra-faint LAEs at $z \sim 3$.

We note that these absorption systems are potentially our first direct probe of the `typical' galaxies that are responsible for hydrogen reionization.  It is widely believed that in order for reionization to complete by $z = 6$, the majority of ionizing photons must come from galaxies below the limits of current imaging surveys \citep[e.g.,][]{bouwens2007,bolton2007b,ouchi2009,bunker2010,mclure2010,oesch2010}.  Metal absorption lines may therefore be the most efficient means of detecting these sources, as directly detecting them in emission must await facilities such as ALMA and JWST.  This study already delivers insights into the properties of reionization galaxies, including the suggestion that they are likely to be more metal poor than their lower-redshift counterparts, at least in the gas phase.  If their stars also have low metallicities, then they may be highly efficient in producing ionizing photons \citep{schaerer2003,venkatesan2003}.  As noted by others \citep[e.g.,][]{ouchi2009}, this would potentially help to explain how reionization is completed by $z \sim 6$.  We add that, in addition to their role in reionization, these low-mass galaxies may contribute significantly to the chemical enrichment of the IGM \citep[e.g.,][]{booth2010}.

\section{Summary}\label{sec:summary}

We have conducted a search for low-ionization metal absorption systems spanning $5.3 < z < 6.4$.  Our survey includes high- and moderate-resolution spectra of 17 QSOs with emission redshifts $z_{\rm em} = 5.8-6.4$.  The total survey pathlength is $\Delta X = 39.5$ ($\Delta z = 8.0$, or $\Delta l = 3.6$~comoving Gpc), of which roughly half is covered by high-resolution data.  We searched for low-ionization systems by looking for coincidences in redshift between \siii, \cii, and \oi.  In total we detect ten systems, five of which were previously reported by \citet{becker2006}.  Each contains \siii\ and \cii, and all but one contains \oi.  The majority are detected in our high-resolution data, which is consistent with the fact that these data are more sensitive to narrow absorption lines.

The line-of-sight number density of absorption systems, uncorrected for completeness, is $\ell(X) = 0.25^{+0.21}_{-0.13}$  (95\% confidence).  This is similar to the number density over $3 < z < 5$ of DLAs and sub-DLAs ($\log{N({\mbox \hi})} \ge 19.0$), which constitute the main population of low-ionization absorbers at those redshifts.  The fact that the number density of low-ionization absorbers is roughly constant out to $z \sim 6$ is in sharp contrast with the evolution of high-ionization systems traced by \civ, which show a marked decline at $z > 5.3$ \citep[Paper I;][]{rw2009}.  At $z \sim 6$, low-ionization systems with $\log{N({\mbox \cii})} \gtrsim 13$ are more common than high-ionization systems with $\log{N({\mbox \civ})} \gtrsim 13$, a reversal from lower redshifts.  

The roughly constant number density of low-ionization systems over $3 \lesssim z \lesssim 6$ may be explained if they are hosted by lower-mass dark matter halos at higher redshifts.  Alternatively, if the systems at $z \sim 6$ are hosted by halos with masses similar to those that host DLAs and sub-DLAs at lower redshifts, then the apparent lack of evolution may occur if dark matter halos of a given mass have a larger mean cross-section of  low-ionization gas at higher redshifts due to the higher gas densities and weaker UV background.

Although the \hi\ column densities cannot be measured for these systems, we are able to infer some of their properties by comparing the velocity widths and metal line strengths to samples at lower redshifts.  For the seven systems with high resolution data, the velocity widths span a similar range as sub-DLAs at $2 \lesssim z \lesssim 4$, although there is some indication that the $z \sim 6$ systems tend to be narrower.  The lines in the $z \sim 6$ systems are also weaker in the sense that, at a given velocity width, the inferred equivalent width of \siii~$\lambda 1526$ is lower than for $z \lesssim 4$ sub-DLAs by a factor of two, and by a factor of six compared to $z \lesssim 4$ DLAs.  This implies that the mass-metallicity relation of the host galaxies evolves towards lower metallicities at higher redshifts, a trend that has also been noted over $2 \lesssim z \lesssim 4$ \citep{ledoux2006}.  Assuming the $z \sim 6$ systems span a similar range in $\log{N({\mbox \hi})}$ as the low-ionization systems at $z  < 4$, the strength of the metal lines implies a decline in the gas-phase metallicity from $2 \lesssim z \lesssim 4$ to $z \sim 6$ of at least $\sim$0.4 dex.  

The mean relative abundances are $[{\rm Si/O}] = -0.03 \pm 0.03$ and $[{\rm C/O}] = -0.18 \pm 0.03$ (1$\sigma$ uncertainties), assuming no dust depletion and no ionization corrections.  These are consistent with the values found for metal-poor DLAs by \citet{pettini2008} and for low-metallicity halo stars by \citet{akerman2004}.  As noted by \citet{becker2006}, the relative abundances are broadly consistent with enrichment from Type II supernovae of low-metallicity massive stars.

Our $z \sim 6$ systems are also notable in that they lack strong high-ionization lines (\siiv\ and \civ), which are ubiquitous among lower-redshift DLAs and sub-DLAs \citep{fox2007a}.  The absence of these lines is consistent with a similar fractional decline in the metallicity of the low- and high-ionization phases, however, and does not necessarily indicate that these systems lack halos of highly-ionized gas.  Deeper spectra will help to determine whether the $z \sim 6$ absorbers have significantly lower $N({\mbox \siiv})/N({\mbox \siii})$ and $N({\mbox \civ})/N({\mbox \cii})$ ratios than lower-redshift systems.

The overall consistency between the properties of the $z \sim 6$ systems and those of lower-redshift DLAs and sub-DLAs suggests that the $z \sim 6$ absorbers arise from galaxy halos, rather than from remnants of neutral IGM at the tail end of hydrogen reionization.  One notable aspect of our survey, however, is that all ten of our systems occur at $z > 5.75$, while roughly 40\% of our pathlength is at lower redshifts.  Expanded surveys over $4.5 < z < 5.7$ will help to determine whether the number density of low-ionization systems truly remains constant out to $z \sim 6$, or whether there is a decline up to $z \sim 5.7$ followed by an increase at higher redshifts.  The latter scenario would potentially indicate a strong evolution in the UV background near $z \sim 6$.

Finally, the high-number density of low-ionization systems at $z \sim 6$ suggests that we are probing galaxies below the detection limits of current $i$-dropout and \lya-emission galaxy surveys.  As such, these absorption systems are potentially the first observations of `typical' galaxies responsible for hydrogen reionization.  The low metallicities we infer suggest that these galaxies may be highly efficient at producing ionizing radiation, a fact which would help to explain how the IGM becomes fully ionized by $z \sim 6$.  As more and higher-redshift QSOs are discovered, absorption lines will continue to provide a unique probe of the reionization era that will complement studies with ALMA, JWST, and other next-generation facilities.

\acknowledgments

We would like to thank Bob Carswell, Andrew Fox,  Max Pettini, and Andrew Pontzen for many helpful discussions during the course of this work.  We also wish to recognize and acknowledge the very significant cultural role and reverence that the summit of Mauna Kea has always had within the indigenous Hawaiian community.  We are most fortunate to have the opportunity to conduct observations from this mountain.  GB has been supported by the Kavli Foundation.  WS received support from the National Science Foundation through grant AST 06-06868.  MR received support from the National Science Foundation through grant AST 05-06845.

\bibliographystyle{apj}
\bibliography{/Users/gdb/tex/refs}

\end{document}